\documentclass[proof]{WileyASNA-v1}

\articletype{Original Paper}%

\received{<day> <Month>, <year>}
\revised{<day> <Month>, <year>}
\accepted{<day> <Month>, <year>}

\begin{document}

\title{Determination of supermassive black hole spins in local active galactic nuclei\\ {\normalsize \it Accepted for publication in Astronomische Nachrichten}}

\author[1,2]{M. Yu. Piotrovich*}

\author[1,2]{S. D. Buliga}

\author[1]{T. M. Natsvlishvili}

\authormark{PIOTROVICH \textsc{et al}}

\address[1]{\orgname{Central Astronomical Observatory at Pulkovo}, \orgaddress{\state{St.-Petersburg}, \country{Russia}}}

\address[2]{\orgname{Special Astrophysical Observatory}, \orgaddress{\state{Nizhnij Arkhyz}, \country{Russia}}}

\corres{*M.Yu. Piotrovich, Central Astronomical Observatory at Pulkovo, St.-Petersburg, Russia. \email{mpiotrovich@mail.ru}}

\abstract{We estimated the radiative efficiency and spin value for a number of local active galactic nuclei with $z < 0.34$ using 3 popular models connecting the radiative efficiency with such parameters of AGNs as mass of supermassive black hole, angle between the line of sight and the axis of the accretion disk and bolometric luminosity. Analysis of the obtained data shown that the spin value decreases with cosmic time, which is in agreement with results of theoretical calculations for low redshift AGNs of other authors. Also we found that the spin value increases with the increasing mass of SMBH and bolometric luminosity. This is the expected result that corresponds to theoretical calculations. Analysis of the distribution of the spin values shown a pronounced peak in the distribution in $0.75 < a < 1.0$ range. $\sim$40\% of objects have spin $a > 0.75$ and $\sim$50\% of objects have spin $a > 0.5$. This results are in a good agreement with our previous results and with the results of other authors.}


\jnlcitation{\cname{%
\author{M.Yu. Piotrovich},
\author{S.D. Buliga}, and
\author{T.M. Natsvlishvili}} (\cyear{2021}),
\ctitle{Determination of supermassive black hole spins in local active galactic nuclei}, \cjournal{Astronomische Nachrichten}, \cvol{2021;00:1--6}.}

\maketitle

\section{Introduction}

Determination of the spin (dimensionless angular momentum) $a = c J / G M_\text{BH}^2$ (where $J$ is the angular momentum, $M_\text{BH}$ is the mass of the black hole and $c$ is the speed of light) of a supermassive black hole (SMBH) located in the center of the active galactic nucleus (AGN) is one of the important problems of modern astrophysics. It has been reliably established that the spin value plays a key role in the generation of relativistic jets in AGNs; therefore, it is the power of the relativistic jet that is most often used to determine the spin of the SMBH \citep{daly11}. As a rule, the kinetic power of a relativistic jet is obtained by estimating the magnetic field strength near the SMBH event horizon using the Blandford–Znajek generation mechanism \citep{blandford77}. Other frequently used mechanisms are the Blandford–Payne mechanism \citep{blandford82} and Garofalo mechanism \citep{garofalo10}.

One of the effective methods for obtaining the spin value $a$ is to determine the radiative efficiency $\varepsilon(a)$ of the accretion disk, which depends significantly on the spin value of the black hole \citep{bardeen72,novikov73,krolik07,krolik07b}. And, in particular, the maximum possible value is $\varepsilon \approx 0.324 $ (at $a \approx 0.998$) \citep{thorne74}. Radiative efficiency is defined as $\varepsilon = L_\text{bol} / \dot{M} c^2$, where $L_{bol}$ is the bolometric luminosity of AGN and $\dot{M}$ is the accretion rate.

\begin{figure*}[!htbp]
\centering
\includegraphics[bb= 75 15 740 525, clip, width=0.67 \columnwidth]{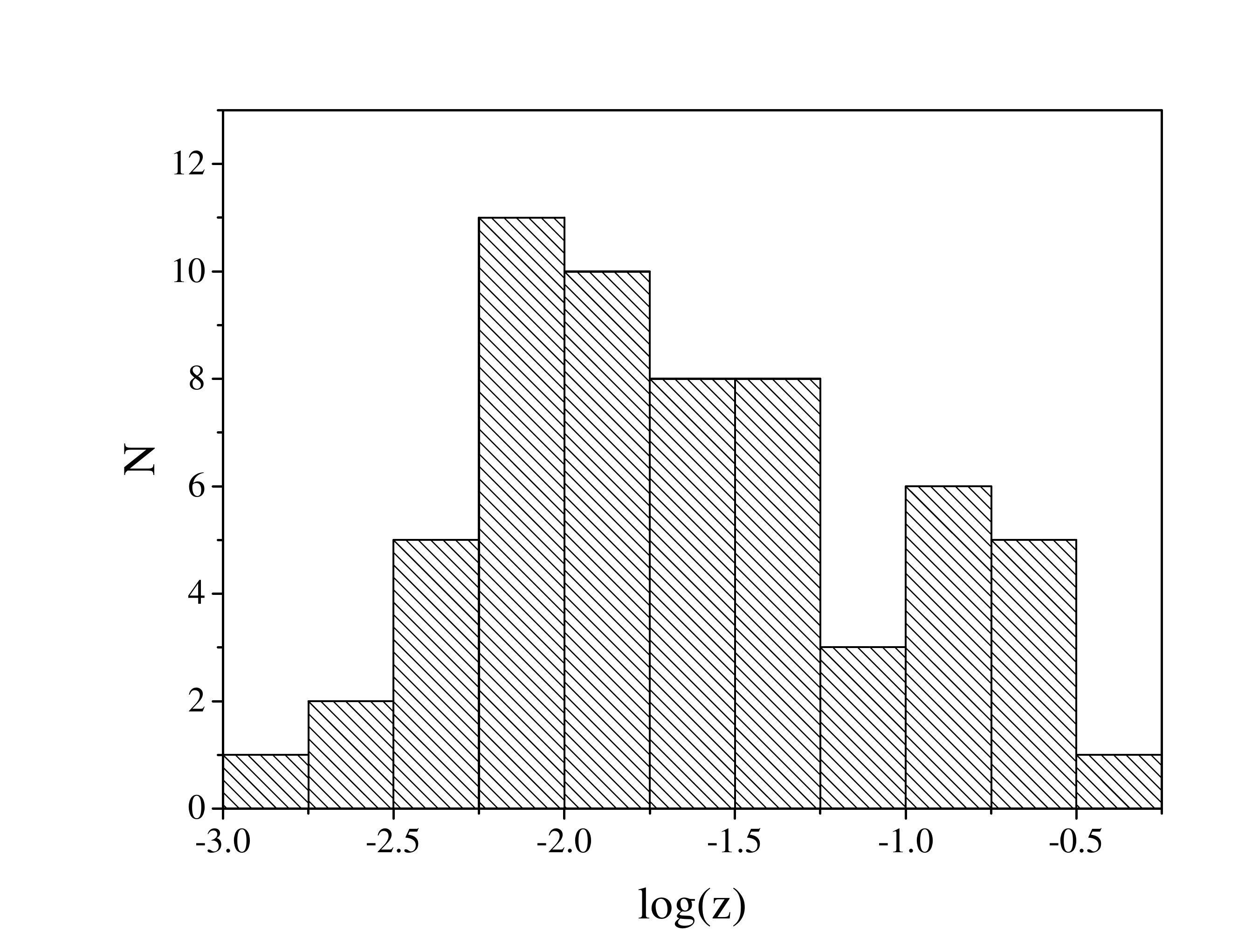}
\includegraphics[bb= 75 10 740 530, clip, width=0.67 \columnwidth]{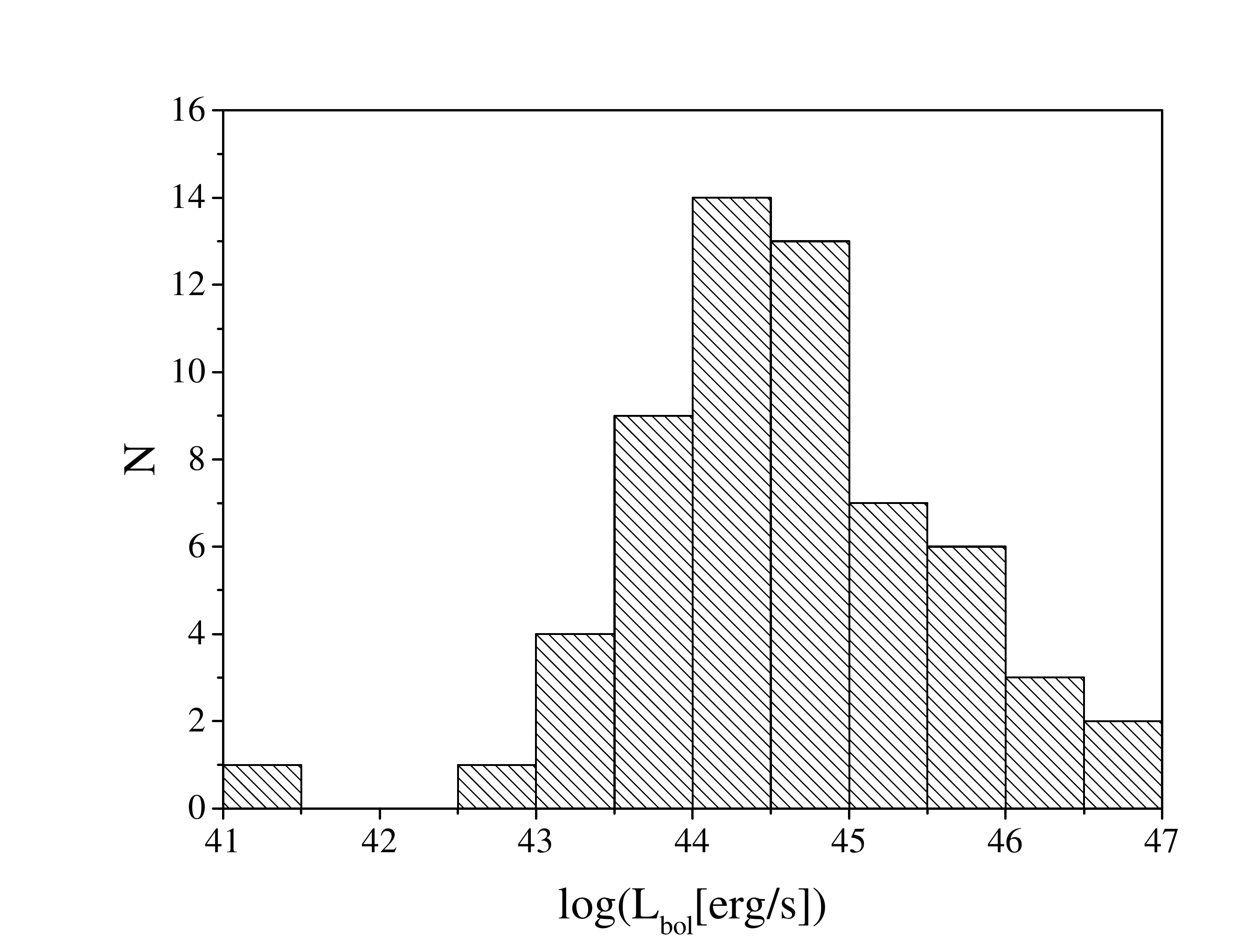}
\includegraphics[bb= 75 10 740 530, clip, width=0.67 \columnwidth]{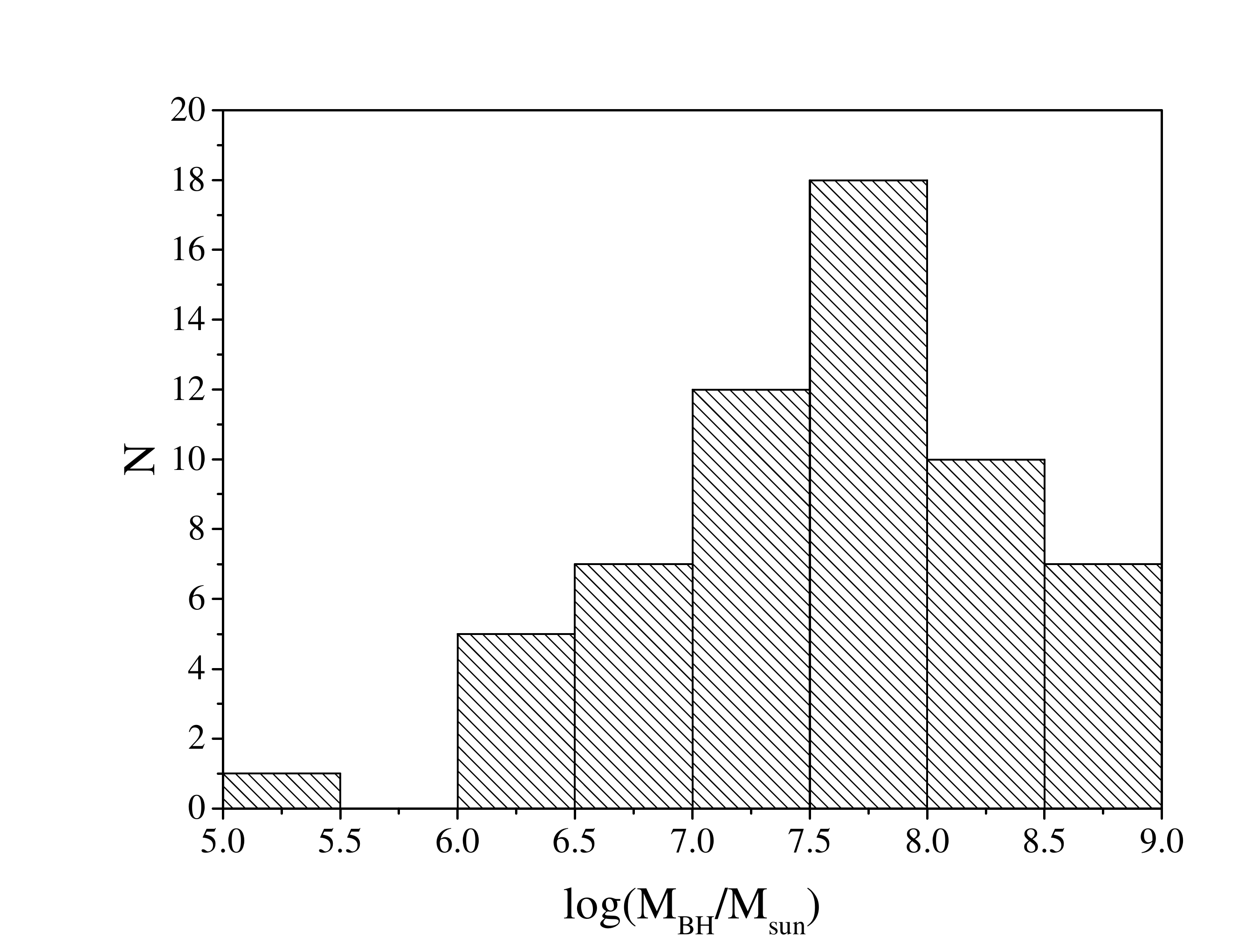}
\caption{Histograms showing the number of objects from our sample with a certain redshift value $z$, bolometric luminosity value $L_\text{bol}$ and SMBH mass value $M_\text{BH}$.}
\label{fig01}
\end{figure*}

\begin{figure*}[!htbp]
\centering
\includegraphics[bb= 65 42 670 530, clip, width=1.0 \columnwidth]{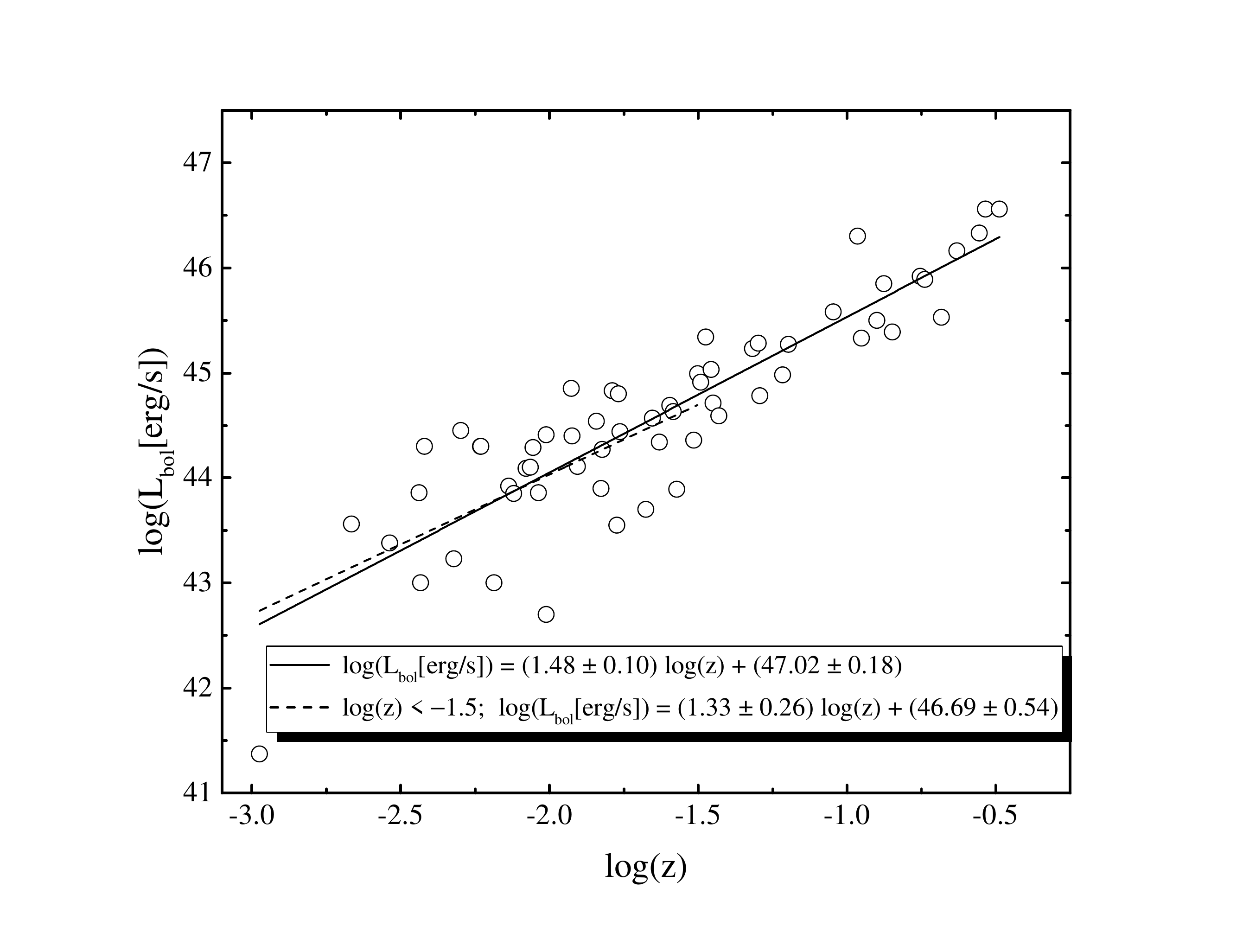}
\includegraphics[bb= 65 40 670 530, clip, width=1.0 \columnwidth]{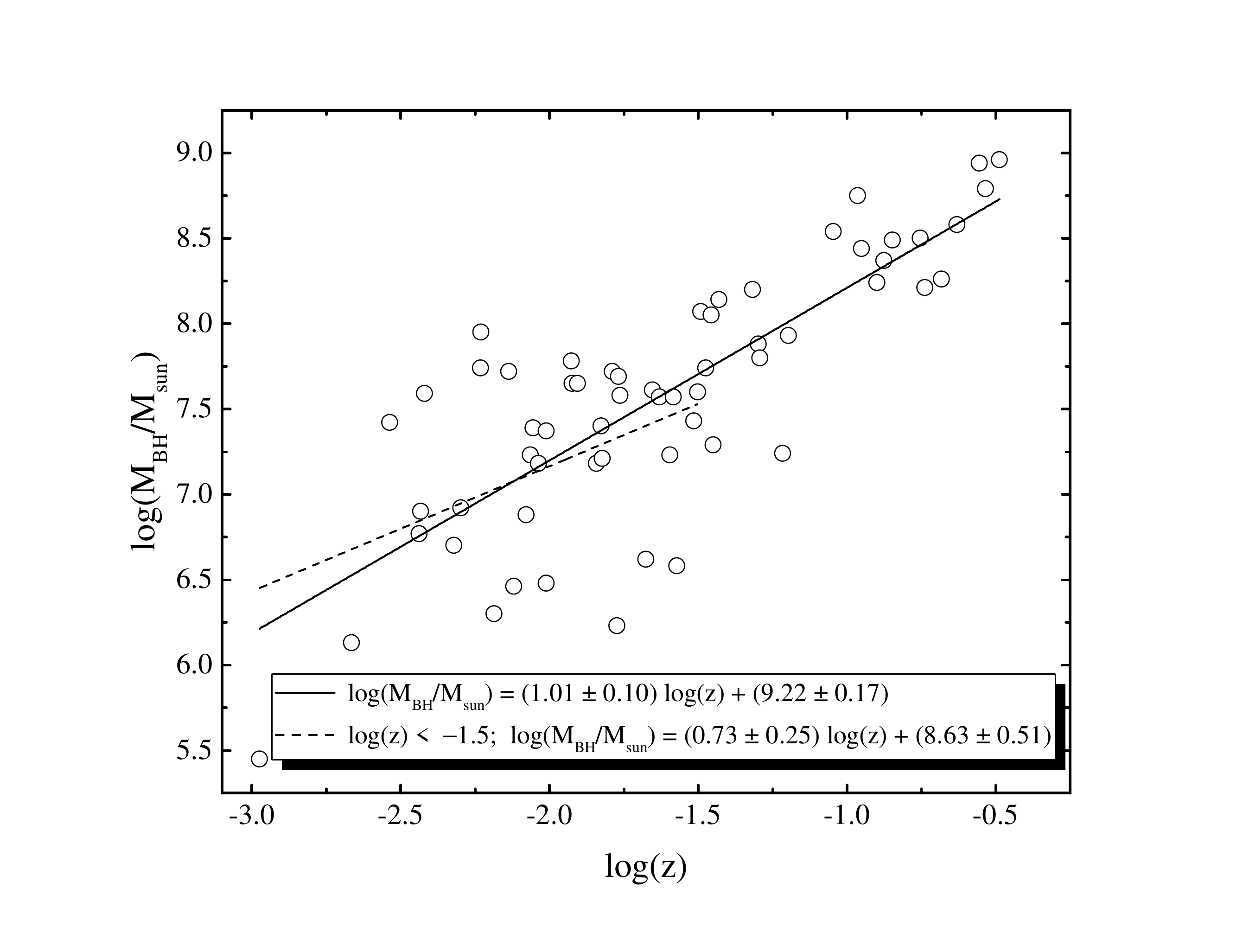}
\caption{Dependences of the bolometric luminosity $L_\text{bol}$ and the SMBH mass $M_\text{BH}$ on the cosmological redshift $z$ for our sample.}
\label{fig02}
\end{figure*}

There are several models (\citet{davis11}, \citet{raimundo11}, \citet{du14}, \citet{trakhtenbrot14}, \citet{lawther17}) connecting the radiative efficiency with such parameters of AGNs as mass of SMBH $M_\text{BH}$, angle between the line of sight and the axis of the accretion disk $i$ and bolometric luminosity $L_\text{bol}$, which could be obtained from observations. The goal of our work is to estimate spin values of SMBHs in AGNs based on this models and the available observational data.

In our work we are using the sample consisted of 111 Seyfert galaxies of the first, second and intermediate types from \citet{marin16}. We choose the objects from \citet{marin16} because for all of this objects Shakura-Sunyaev model \citep{shakura73} of a geometrically thin, optically thick accretion disk can be used. And all methods for calculating radiative efficiency used in our paper assume this disk model.

\begin{table*}
\caption{Estimated values of spin for all models: $a_1$ - \citet{raimundo11}, $a_2$ - \citet{trakhtenbrot14}, $a_3$ - \citet{du14}, $L_{bol}$ in [erg/s], $i$ in [deg].}
\centering
\footnotesize
\begin{tabular}{lccccccccc}
\hline
Object & Type & $\log(z)$ & $\log{\frac{M_{BH}}{M_{\odot}}}$ & $\log{L_{bol}}$ & $\log{l_E}$ & $i$ & $a_1$ & $a_2$ & $a_3$\\
\hline
3C 120 & S1.5 & -1.474 & 7.74 & 45.34 & -0.58 & 22.0 & 0.332 & 0.690 & 0.768 \\
Akn 120 & BLS1 & -1.491 & 8.07 & 44.91 & -1.34 & 42.0 & 0.946 & 0.996 & 0.996 \\
Arp 151 & BLS1 & -1.676 & 6.62 & 43.70 & -1.10 & 25.2 & 0.524 & 0.558 & -0.070 \\
ESO 323-G077 & NLS1 & -1.827 & 7.40 & 43.90 & -1.68 & 45.0 & 0.964 & 0.990 & 0.944 \\
ESO 362-G18 & S1.5 & -1.905 & 7.65 & 44.11 & -1.72 & 53.0 & 0.952 & 0.990 & 0.960 \\
Fairall 9 & BLS1 & -1.317 & 8.20 & 45.23 & -1.15 & 35.0 & 0.936 & 0.996 & 0.998 \\
I Zw1 & NLS1 & -1.216 & 7.24 & 44.98 & -0.44 & 8.0 & -0.402 & 0.026 & 0.084 \\
IC 2560 & S2.0 & -2.011 & 6.48 & 42.70 & -1.96 & 66.0 & 0.778 & 0.768 & -- \\
IRAS 13349+2438 & S2.0 & -0.964 & 8.75 & 46.30 & -0.63 & 52.0 & 0.524 & 0.910 & 0.968 \\
K 348-7 & S1.0 & -0.631 & 8.58 & 46.16 & -0.60 & 35.0 & 0.736 & 0.958 & 0.990 \\
LEDA 46718 & S1.5 & -1.767 & 7.69 & 44.80 & -1.07 & 27.0 & 0.836 & 0.944 & 0.938 \\
LEDA 47969 & S1.5 & -2.120 & 6.46 & 43.85 & -0.79 & 34.0 & -0.754 & -0.754 & -- \\
Mrk 110 & S1.5 & -1.450 & 7.29 & 44.71 & -0.76 & 37.4 & -0.070 & 0.314 & 0.232 \\
Mrk 279 & BLS1 & -1.514 & 7.43 & 44.36 & -1.25 & 35.0 & 0.836 & 0.926 & 0.874 \\
Mrk 335 & NLS1 & -1.595 & 7.23 & 44.69 & -0.72 & 20.0 & 0.210 & 0.474 & 0.416 \\
Mrk 34 & S2.0 & -1.293 & 7.80 & 44.78 & -1.20 & 65.0 & -0.070 & 0.460 & 0.430 \\
Mrk 348 & S2.0 & -1.823 & 7.21 & 44.27 & -1.12 & 60.0 & -0.506 & -0.038 & -0.452 \\
Mrk 359 & NLS1 & -1.774 & 6.23 & 43.55 & -0.86 & 30.0 & -0.506 & -0.624 & -- \\
Mrk 463 & S2.0 & -1.298 & 7.88 & 45.28 & -0.78 & 60.0 & -0.824 & 0.084 & 0.254 \\
Mrk 50 & S1.0 & -1.630 & 7.57 & 44.34 & -1.41 & 9.0 & 0.984 & 1.000 & 0.992 \\
Mrk 509 & BLS1 & -1.456 & 8.05 & 45.03 & -1.20 & 19.0 & 0.970 & 1.000 & -- \\
Mrk 573 & S2.0 & -1.762 & 7.58 & 44.44 & -1.32 & 60.0 & 0.400 & 0.682 & 0.568 \\
Mrk 590 & BLS1 & -1.584 & 7.57 & 44.63 & -1.12 & 17.8 & 0.876 & 0.956 & 0.944 \\
Mrk 78 & S2.0 & -1.430 & 8.14 & 44.59 & -1.73 & 60.0 & 0.950 & 0.996 & 0.992 \\
Mrk 79 & BLS1 & -1.654 & 7.61 & 44.57 & -1.22 & 58.0 & 0.332 & 0.660 & 0.580 \\
Mrk 817 & S1.5 & -1.502 & 7.60 & 44.99 & -0.79 & 41.6 & 0.162 & 0.558 & 0.590 \\
Mrk 877 & BLS1 & -0.951 & 8.44 & 45.33 & -1.29 & 20.0 & 0.998 & -- & -- \\
Mrk 896 & NLS1 & -1.572 & 6.58 & 43.89 & -0.87 & 15.0 & 0.136 & 0.186 & -0.564 \\
NGC 1068 & S2.0 & -2.419 & 7.59 & 44.30 & -1.47 & 70.0 & -0.106 & 0.350 & 0.084 \\
NGC 1320 & S2.0 & -2.036 & 7.18 & 43.86 & -1.50 & 68.0 & -0.106 & 0.210 & -0.564 \\
NGC 1386 & S1.9/2.0 & -2.537 & 7.42 & 43.38 & -2.22 & 81.0 & -0.106 & 0.274 & -- \\
NGC 1566 & S1.5 & -2.298 & 6.92 & 44.45 & -0.65 & 30.0 & -0.624 & -0.262 & -0.624 \\
NGC 2992 & S2.0 & -2.137 & 7.72 & 43.92 & -1.98 & 70.0 & 0.828 & 0.938 & 0.816 \\
NGC 3227 & S1.5 & -2.438 & 6.77 & 43.86 & -1.09 & 14.2 & 0.652 & 0.710 & 0.350 \\
NGC 3516 & S1.5 & -2.055 & 7.39 & 44.29 & -1.28 & 26.0 & 0.900 & 0.960 & 0.916 \\
NGC 3783 & S1.5 & -2.011 & 7.37 & 44.41 & -1.14 & 15.0 & 0.854 & 0.934 & 0.894 \\
NGC 4051 & NLS1 & -2.666 & 6.13 & 43.56 & -0.75 & 19.6 & -0.754 & -- & -- \\
NGC 424 & S2.0 & -1.927 & 7.78 & 44.85 & -1.11 & 69.0 & -- & -0.142 & -0.142 \\
NGC 4388 & S2.0 & -2.064 & 7.23 & 44.10 & -1.31 & 60.0 & 0.136 & 0.416 & 0.026 \\
NGC 4395 & S1.8 & -2.975 & 5.45 & 41.37 & -2.26 & 15.0 & -- & -- & 0.084 \\
NGC 4507 & S1.9/2.0 & -1.924 & 7.65 & 44.40 & -1.43 & 47.0 & 0.882 & 0.962 & 0.932 \\
NGC 4593 & BLS1 & -2.079 & 6.88 & 44.09 & -0.97 & 21.6 & 0.446 & 0.568 & 0.254 \\
NGC 4941 & S2.0 & -2.433 & 6.90 & 43.00 & -2.08 & 70.0 & 0.796 & 0.842 & -0.070 \\
NGC 5506 & NLS1 & -2.230 & 7.95 & 44.30 & -1.83 & 80.0 & -0.824 & 0.084 & -0.262 \\
NGC 5548 & S1.5 & -1.789 & 7.72 & 44.83 & -1.07 & 47.3 & 0.568 & 0.812 & 0.804 \\
NGC 7213 & LINER & -2.231 & 7.74 & 44.30 & -1.62 & 21.0 & 1.000 & -- & -- \\
NGC 7314 & S1.9 & -2.321 & 6.70 & 43.23 & -1.65 & 42.0 & 0.922 & 0.936 & 0.536 \\
PG 0026+129 & NLS1 & -0.848 & 8.49 & 45.39 & -1.28 & 43.0 & 0.972 & -- & -- \\
PG 1302-102 & S1.0 & -0.555 & 8.94 & 46.33 & -0.79 & 32.0 & 0.954 & -- & -- \\
PG 1411+442 & BLS1 & -1.046 & 8.54 & 45.58 & -1.14 & 14.0 & 0.994 & -- & -- \\
PG 1435-067 & BLS1 & -0.900 & 8.24 & 45.50 & -0.92 & 38.0 & 0.808 & 0.964 & 0.982 \\
PG 1626+554 & BLS1 & -0.876 & 8.37 & 45.85 & -0.70 & 31.0 & 0.768 & 0.956 & 0.984 \\
PG 1700+518 & NLS1 & -0.535 & 8.79 & 46.56 & -0.41 & 43.0 & 0.524 & 0.914 & 0.974 \\
PG 2251+113 & S1.0 & -0.488 & 8.96 & 46.56 & -0.58 & 67.0 & -0.452 & 0.626 & 0.824 \\
RBS 1124 & BLS1 & -0.682 & 8.26 & 45.53 & -0.91 & 66.0 & -0.452 & 0.416 & 0.568 \\
Swift J2127.4+5654 & NLS1 & -1.842 & 7.18 & 44.54 & -0.82 & 49.0 & -0.686 & -0.180 & -0.402 \\
Ton 1388 & S1.0 & -0.753 & 8.50 & 45.92 & -0.76 & 39.0 & 0.782 & 0.966 & 0.990 \\
Ton 1542 & BLS1 & -1.197 & 7.93 & 45.27 & -0.84 & 28.0 & 0.724 & 0.912 & 0.938 \\
Ton 1565 & S1.0 & -0.738 & 8.21 & 45.89 & -0.50 & 37.0 & 0.314 & 0.768 & 0.874 \\
UGC 6728 & S1.0 & -2.186 & 6.30 & 43.00 & -1.48 & 55.0 & 0.416 & 0.350 & -- \\
\hline
\end{tabular}
\label{tab1}
\end{table*}

\section{Vetting of the initial data}

First, we will vet the initial data from \citet{marin16}.

Fig.~\ref{fig01} show distribution of objects by the redshift, mass and bolometric luminosity. The redshift distribution shows an increase at small $z$, associated with an increase in the number of objects with increasing distance, and then the number of objects begins to decrease, which is apparently associated with the selection effect, since at large distances the number of objects, which brightness makes it possible to confidently determine their physical parameters, decreases. However, we can see that mass and luminosity distributions have a normal form with a peak in the region of the known mean value of the parameters and thus have no visible manifestations of the selection effect, which could arise because among distant objects we could primarily observe the brightest ones with the largest mass of the central SMBH.

Fig.~\ref{fig02} demonstrate dependences of bolometric luminosity and SMBH mass on redshift. In order to test the possible influence of selection effect on this dependences we make two linear fittings: for all objects and for nearest objects with $\log{z} < -1.5$. For bolometric luminosity we have: $\log{L_\text{bol}\text{[erg/s]}} = (1.48\pm 0.10) \log{z} + (47.02\pm 0.18)$ for all objects and $\log{L_\text{bol}\text{[erg/s]}} = (1.33\pm 0.26) \log{z} + (46.69\pm 0.54)$ for objects with $\log{z} < -1.5$. For SMBH mass we have: $\log{M_\text{BH}/M_\text{sun}} = (1.01\pm 0.10) \log{z} + (9.22\pm 0.17)$ for all objects and $\log{M_\text{BH}/M_\text{sun}} = (0.73\pm 0.25) \log{z} + (8.63\pm 0.51)$ for objects with $\log{z} < -1.5$. In both cases the fittings are close within a margin of error. Thus we can conclude that influence of selection effect is relatively weak.

\section{Estimations of radiative efficiency and spin}

We estimated the value of the radiative efficiency $\varepsilon$ from observational data for a number of local active galaxies. The initial sample consisted of 111 Seyfert galaxies of the first, second and intermediate types from \citet{marin16}. The masses of SMBHs $M_\text{BH}$, the bolometric luminosities $L_\text{bol}$, and the angles $i$ were all taken from \citet{marin16}.

In \citet{marin16} mass and luminosity values were taken from literature and inclination angles were estimated by Marin through several different methods. Such as X-ray spectroscopy, IR spectroscopy, narrow line region (NLR) spectroscopy and other methods. Admittedly, all of these methods have their drawbacks, but at the moment, unfortunately, there is no absolutely reliable method.

There are number of models, that allow us to estimate the radiative efficiency $\varepsilon$ from this parameters (\citet{davis11}, \citet{raimundo11}, \citet{du14}, \citet{trakhtenbrot14}, \citet{lawther17}). To get relationship between parameters all of this methods use statistical analysis of observational data on AGNs and Shakura-Sunyaev accretion disk model \citep{shakura73}. The methods of \citet{davis11} and \citet{raimundo11} are basically the same. The \citet{lawther17} method is a restatement of the \citet{raimundo11} method, where Eq.6 in \citet{raimundo11} assumes that optical luminosity  $L_\text{opt}$ is measured at 4392 \AA, and Eq.4 in \citet{lawther17} contains a wavelength scaling factor. So, the estimations was carried out using the following three different enough models (we changed the form of equations from original papers for greater consistency):

\begin{enumerate}
\item \citet{raimundo11}:\\$\varepsilon \left( a \right) = 0.063\left( {\frac{L_\text{bol} }{10^{46}\text{erg/s}}} \right)^{0.99}\left( {\frac{L_\text{opt} }{10^{45}\text{erg/s}}} \right)^{-1.5} M_8^{0.89} \mu^{1.5}$.

\item \citet{trakhtenbrot14}:\\$\varepsilon \left( a \right) =  0.073 \left(\frac{L_\text{bol}}{10^{46}\text{erg/s}}\right) \left(\frac{\lambda L_\lambda}{10^{45}\text{erg/s}} \right)^{-1.5} \left(\frac{\lambda}{5100\text{\AA}}\right)^{-2} M_8 \mu^{1.5}\\ \lambda L_\lambda = L_{opt}, \lambda = 4400\text{\AA}$.

\item \citet{du14}:\\ $\varepsilon \left( a \right) =  0.105 \left(\frac{L_\text{bol}}{10^{46}\text{erg/s}}\right) \left(\frac{L_{5100}}{10^{45}\text{erg/s}} \right)^{-1.5} M_8 \mu^{1.5}$.
\end{enumerate}

\noindent Here $L_{5100}$ is luminosity at 5100 \AA, $M_8 = M_\text{BH} / (10^8 M_{\odot})$ and $\mu = \cos{i}$. For third model \citep{du14} we use the Eddington ratio $l_\text{E} = L_\text{bol} / L_\text{Edd}$, where $L_\text{Edd} = 1.5 \times 10^{38} M_\text{BH} / M_\odot $ is the Eddington luminosity.

There is serious problem in defining the bolometric correction factors that are used for obtaining luminosity value at a certain wavelength from bolometric luminosity. In the literature, we can find factors that differ 2-3 times \citep{richards06, hopkins07, cheng19, netzer19, duras20}. In this work we decided to use for consistency bolometric correction for luminosity at 5100\AA\, from \citet{richards06}: $L_{5100} = L_\text{bol} / 10.3$ and definition of optical luminosity $L_{opt}$ from \citet{hopkins07}: $L_\text{bol} / L_\text{opt} = L_\text{bol} / L_\text{B} = 6.25 (L_\text{bol} / (10^{10} L_\odot))^{-0.37} + 9.0 (L_\text{bol} / (10^{10} L_\odot))^{-0.012}, L_\text{B} =L(\text{4400\AA})$.

According to \citet{thorne74} radiative efficiency should be in the range $0.039 < \varepsilon(a) < 0.324$. After calculations, we got 68 objects for which at least one of the models gave a result within these range. The spin value $a$ was determined numerically using the relation \citep{bardeen72}:

\begin{equation}
 \varepsilon(a) = 1 - \frac{R_\text{ISCO}^{3/2} - 2 R_\text{ISCO}^{1/2} + |a|}{R_\text{ISCO}^{3/4}(R_\text{ISCO}^{3/2} - 3 R_\text{ISCO}^{1/2} + 2 |a|)^{1/2}},
 \label{eq01}
\end{equation}

\noindent where $R_\text{ISCO}$ is the radius of the innermost stable circular orbit of a black hole, which is expressed through the spin as follows:

\begin{equation}
 \begin{array}{l}
  R_\text{ISCO}(a) = 3 + Z_2 \pm ((3 - Z_1)(3 + Z_1 + 2 Z_2))^{1/2},\\
  Z_1 = 1 + (1 - a^2)^{1/3}((1 + a)^{1/3} + (1 - a)^{1/3}),\\
  Z_2 = (3 a^2 + Z_1^2)^{1/2}.
 \end{array}
 \label{eq02}
\end{equation}

\noindent In the expression for $R_\text{ISCO}(a)$, the sign ''-'' is used for prograde ($a \geq 0$), and the sign ''+'' for retrograde rotation ($a < 0$).

\begin{figure*}[!htbp]
\centering
\includegraphics[bb= 57 30 670 525, clip, width=0.67 \columnwidth]{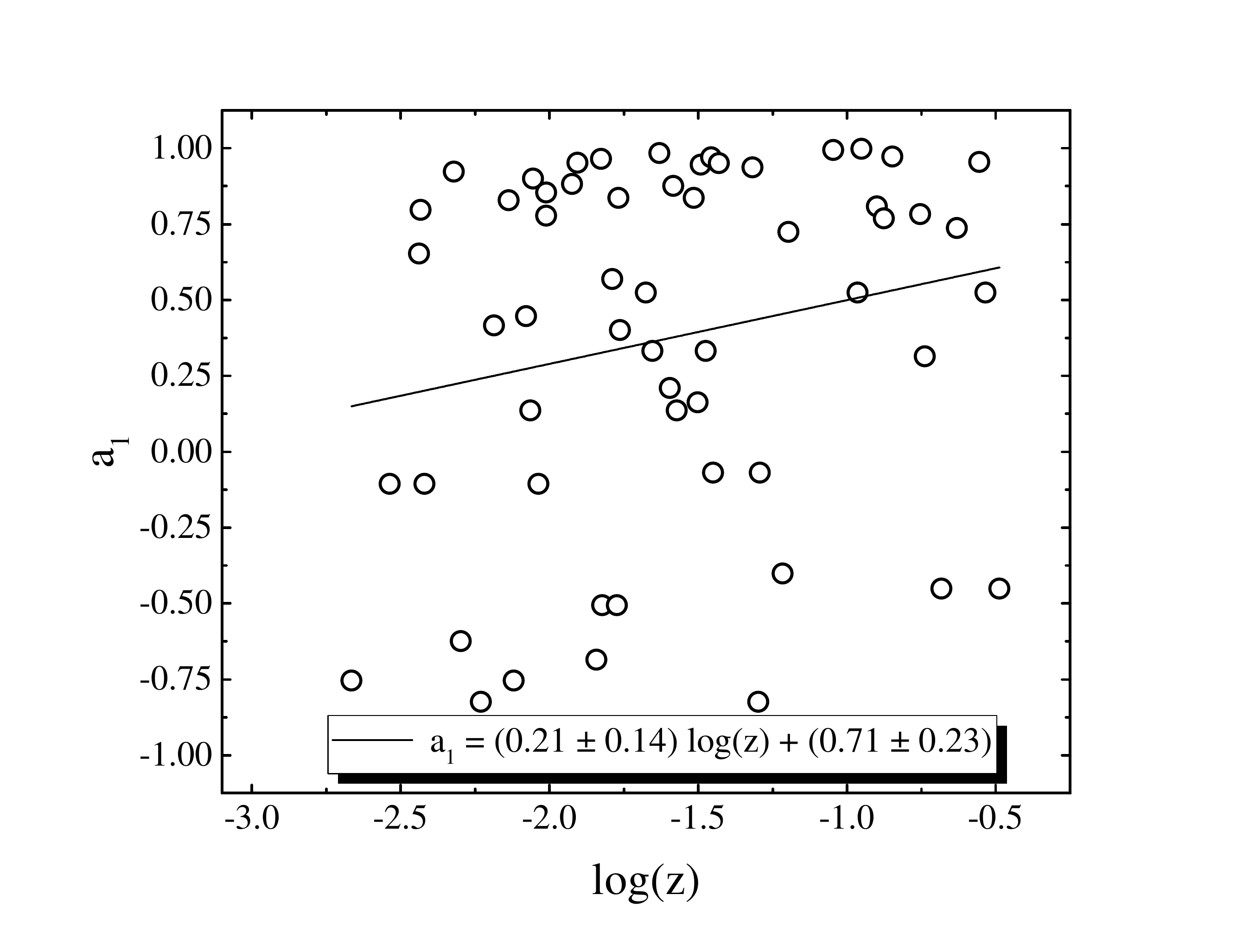}
\includegraphics[bb= 57 25 675 525, clip, width=0.67 \columnwidth]{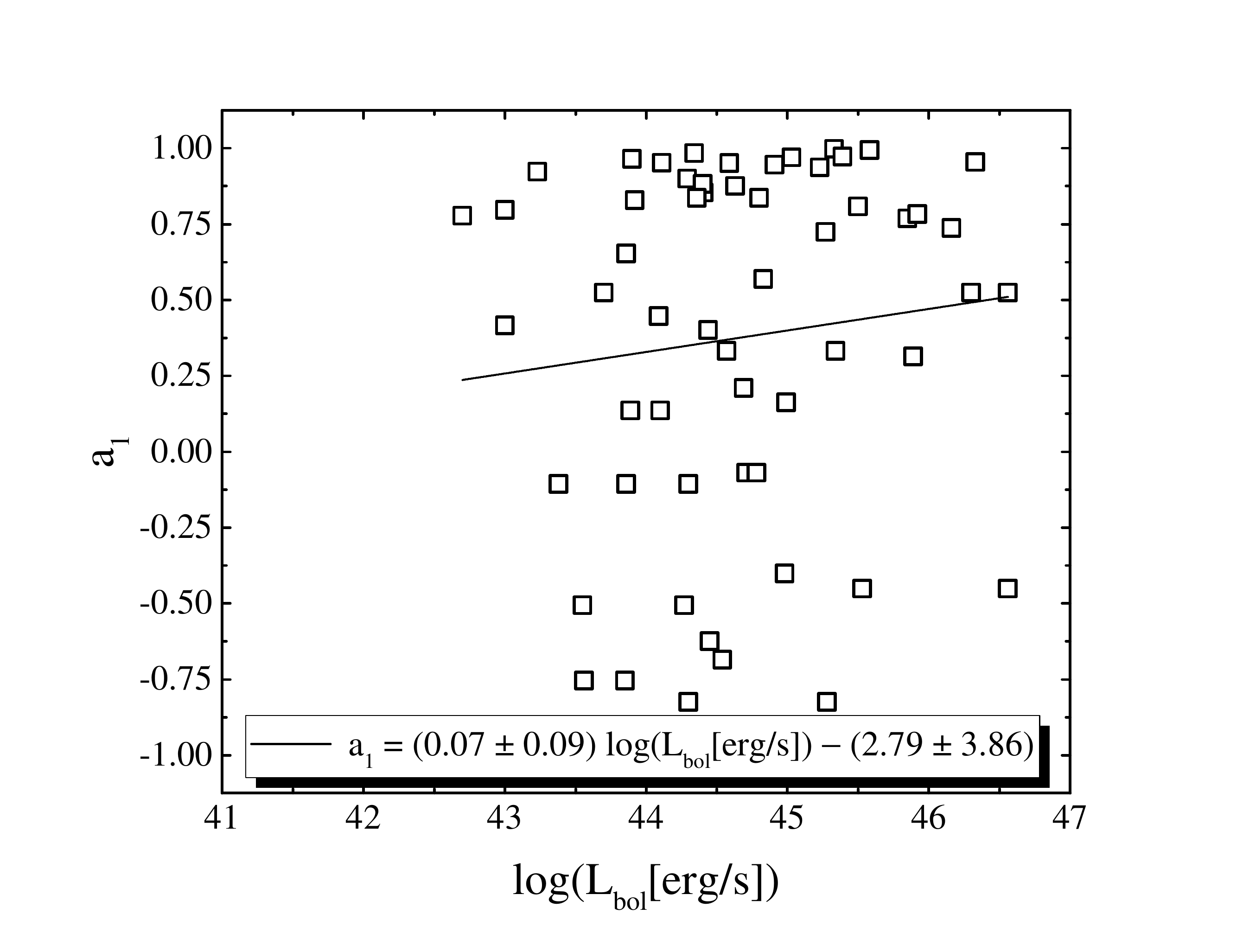}
\includegraphics[bb= 58 25 680 525, clip, width=0.67 \columnwidth]{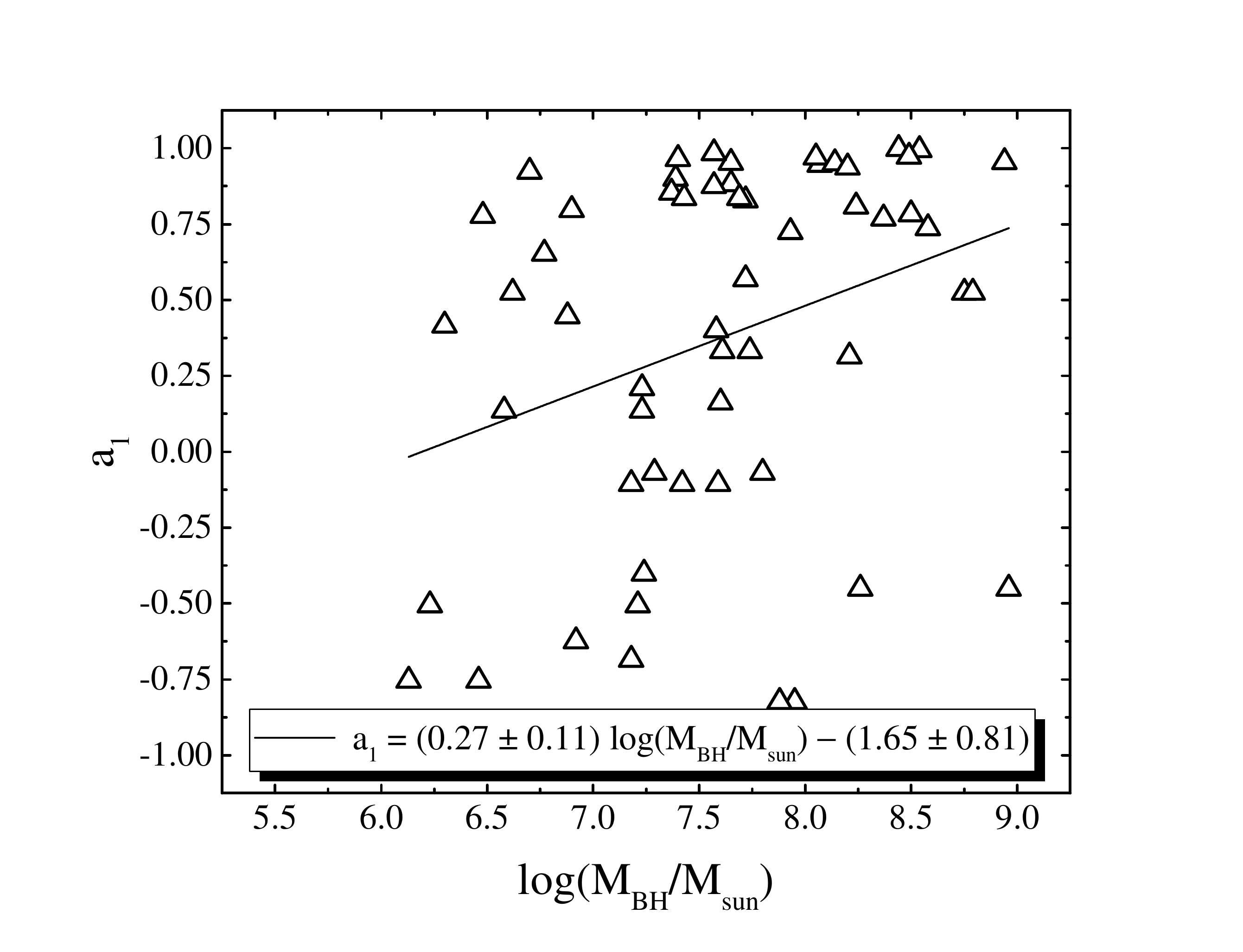}
\caption{Dependence of the spin value $a$ on the redshift $z$, bolometric luminosity $L_\text{bol}$ and the SMBH mass $M_\text{BH}$ for the first model \citep{raimundo11}.}
\label{fig03}
\end{figure*}

\begin{figure*}[!htbp]
\centering
\includegraphics[bb= 57 30 670 525, clip, width=0.67 \columnwidth]{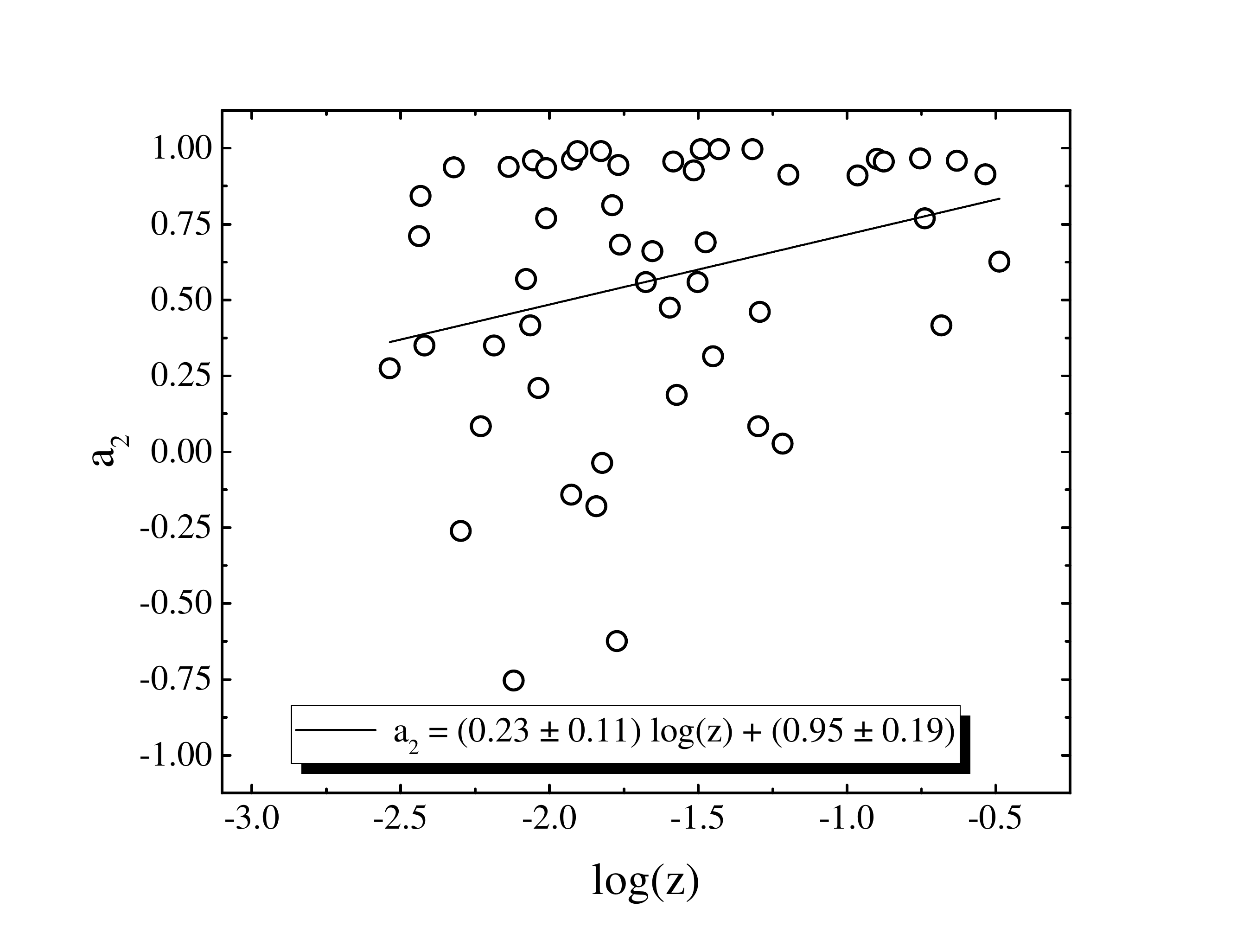}
\includegraphics[bb= 57 25 675 525, clip, width=0.67 \columnwidth]{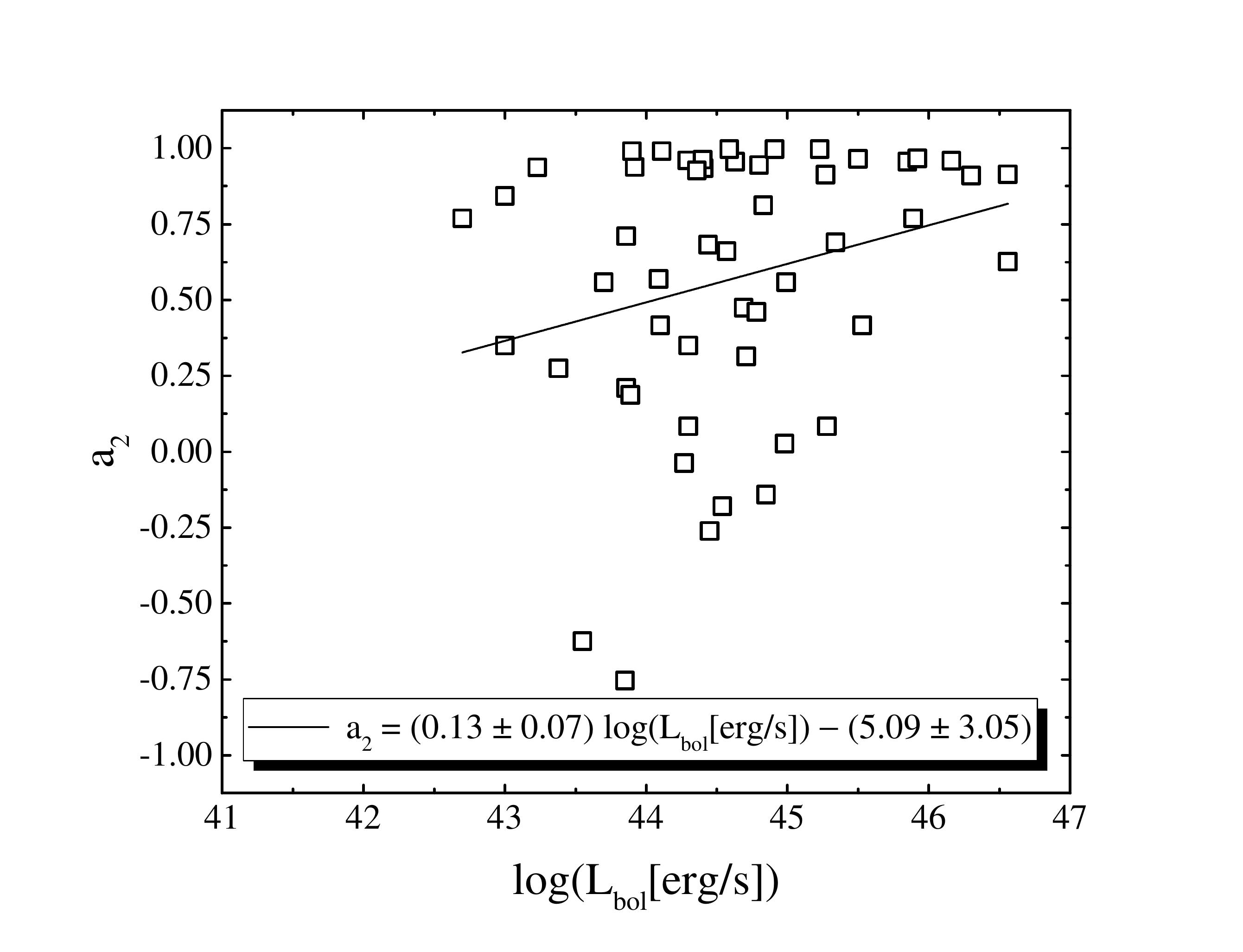}
\includegraphics[bb= 58 25 680 525, clip, width=0.67 \columnwidth]{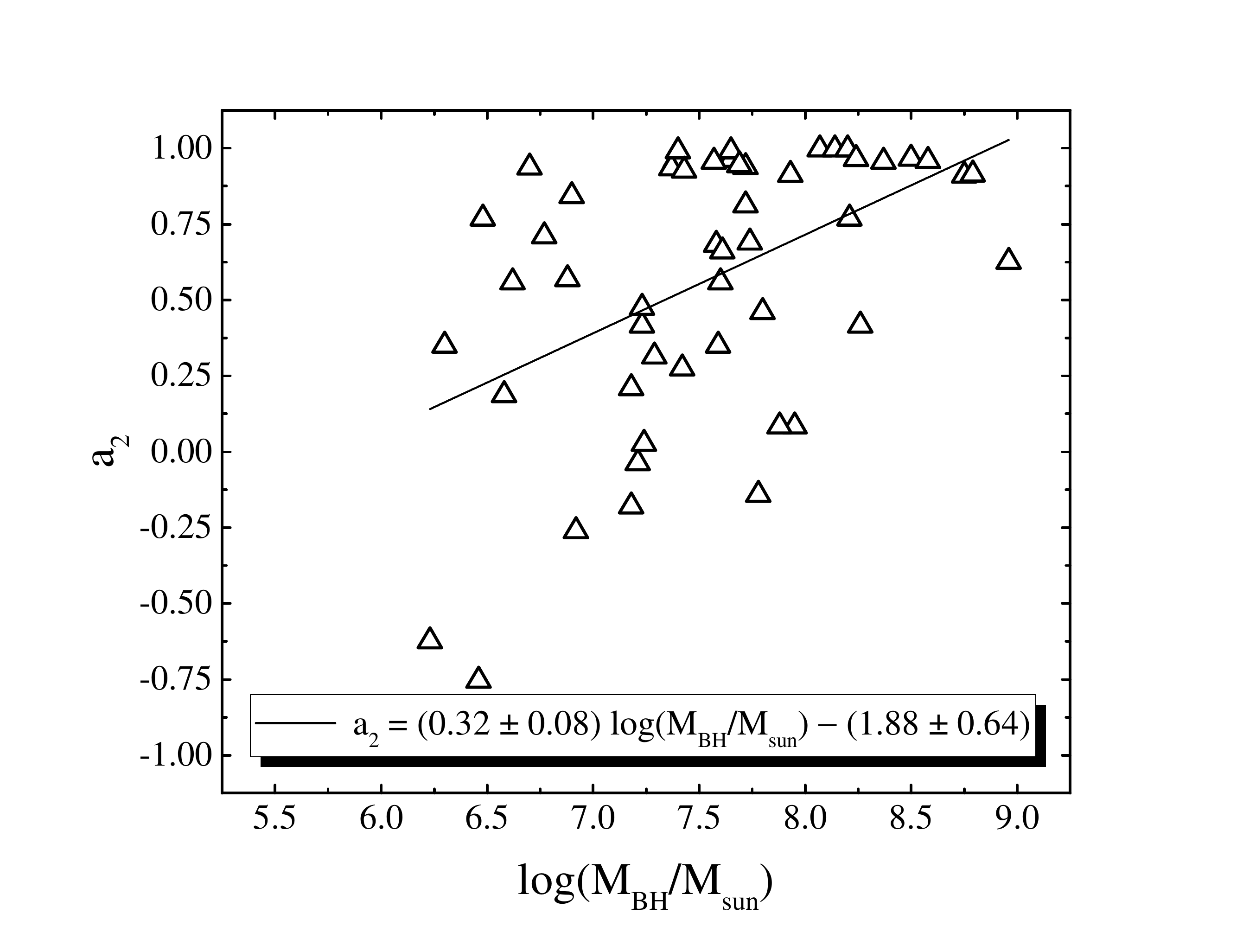}
\caption{Dependence of the spin value $a$ on the redshift $z$, bolometric luminosity $L_\text{bol}$ and the SMBH mass $M_\text{BH}$ for the second model \citep{trakhtenbrot14}.}
\label{fig04}
\end{figure*}

\begin{figure*}[!htbp]
\centering
\includegraphics[bb= 57 30 670 525, clip, width=0.67 \columnwidth]{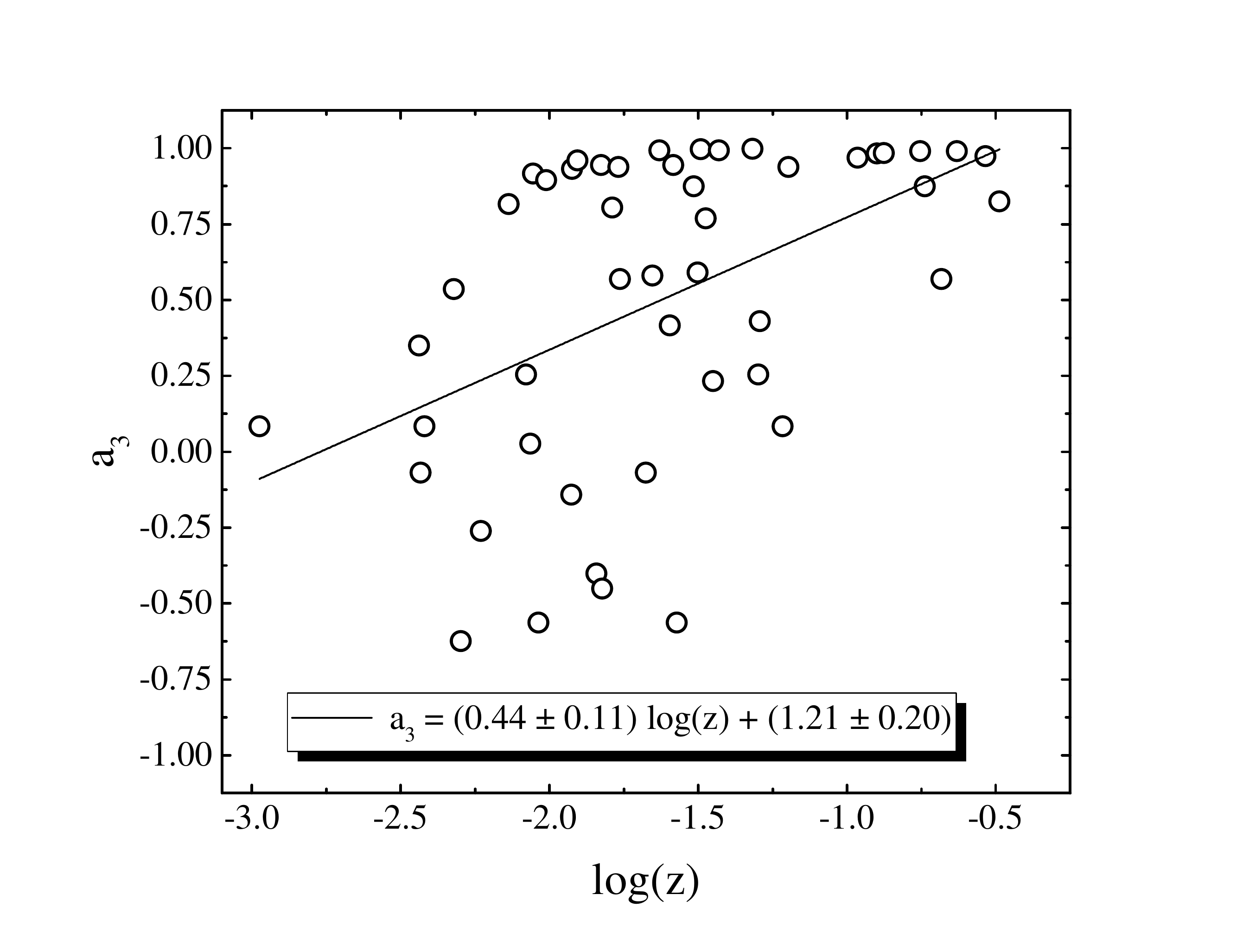}
\includegraphics[bb= 57 25 675 525, clip, width=0.67 \columnwidth]{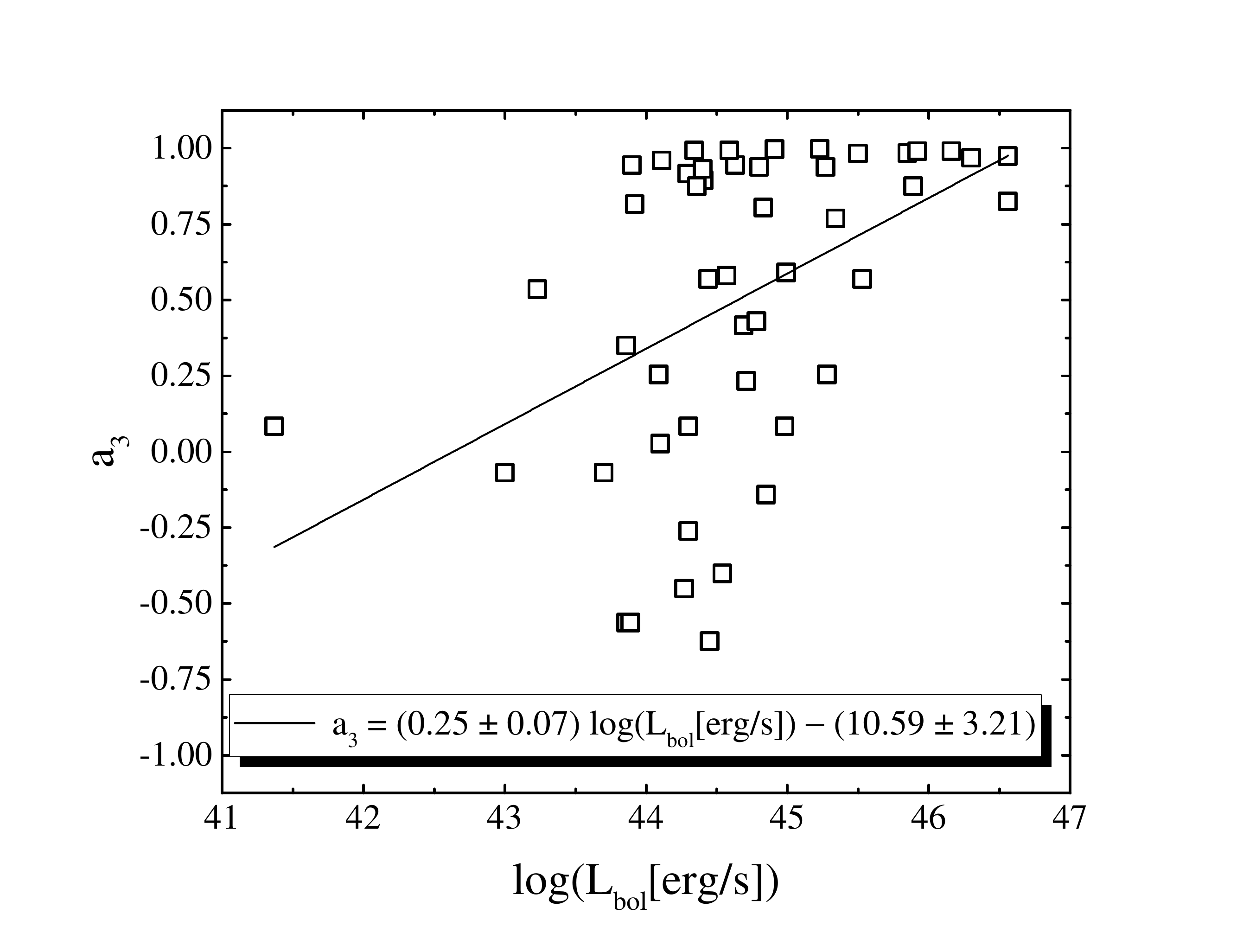}
\includegraphics[bb= 58 25 680 525, clip, width=0.67 \columnwidth]{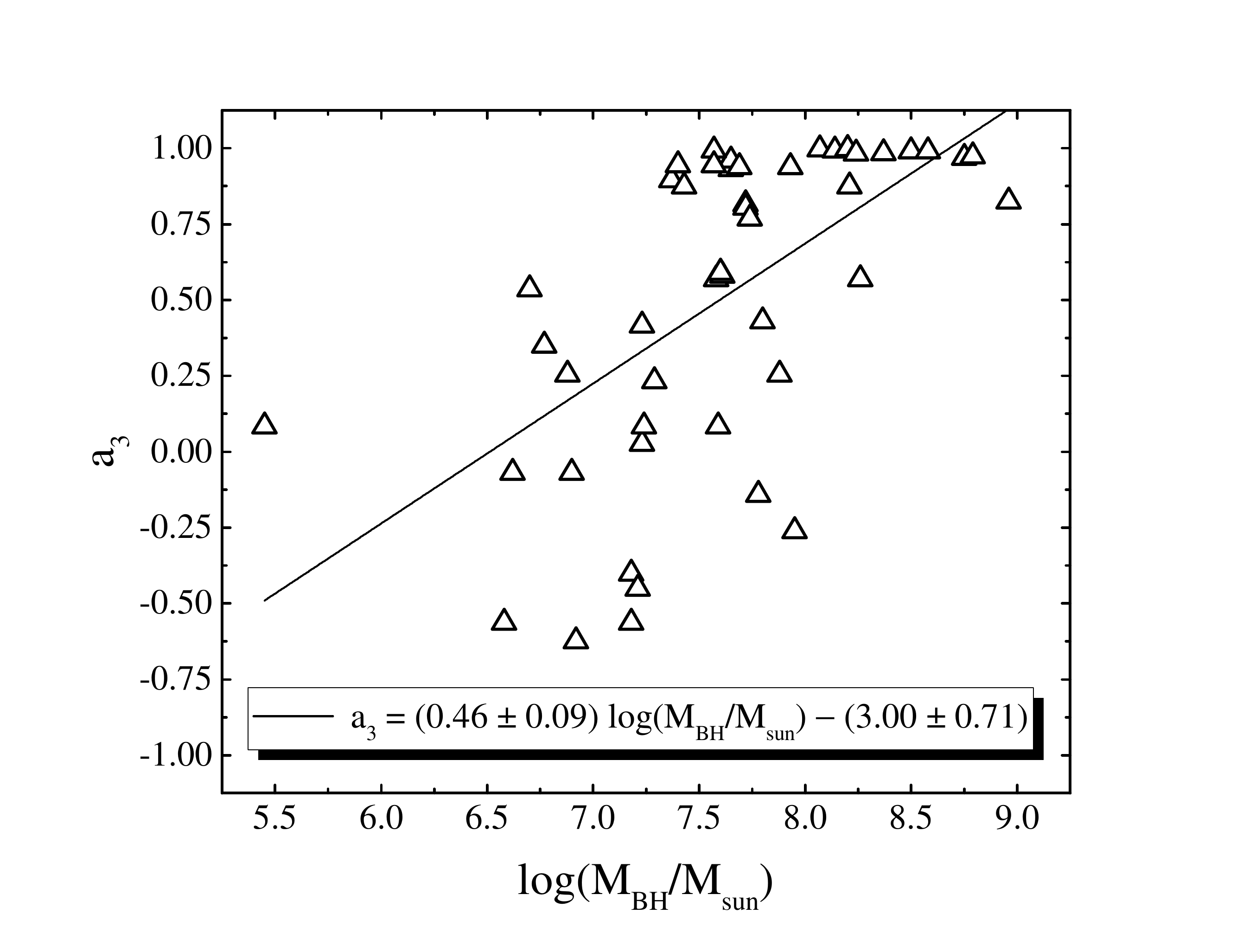}
\caption{Dependence of the spin value $a$ on the redshift $z$, bolometric luminosity $L_\text{bol}$ and the SMBH mass $M_\text{BH}$ for the third model \citep{du14}.}
\label{fig05}
\end{figure*}

Table~\ref{tab1} presents the results of our calculations of spin $a$ for all three models. Sign "--" means that the value is out of range. Table~\ref{tab1} also shows type of object, cosmological redshift $z$, mass of SMBH $M_\text{BH}$, bolometric luminosity $L_\text{bol}$, Eddington ratio $l_\text{E}$ and inclination angle $i$.

\begin{figure*}[!htbp]
\centering
\includegraphics[bb= 58 11 726 525, clip, width=0.67 \columnwidth]{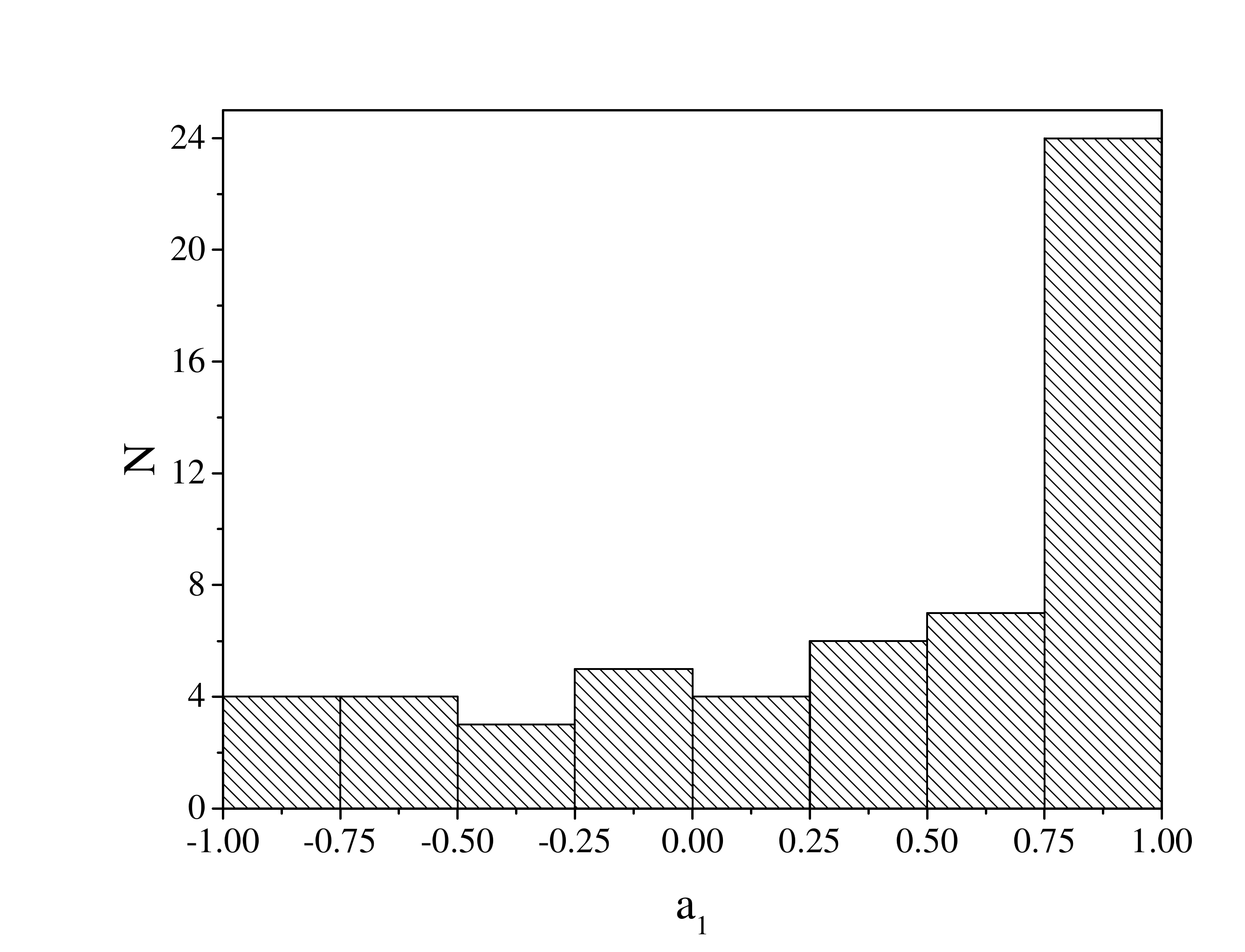}
\includegraphics[bb= 58 11 726 525, clip, width=0.67 \columnwidth]{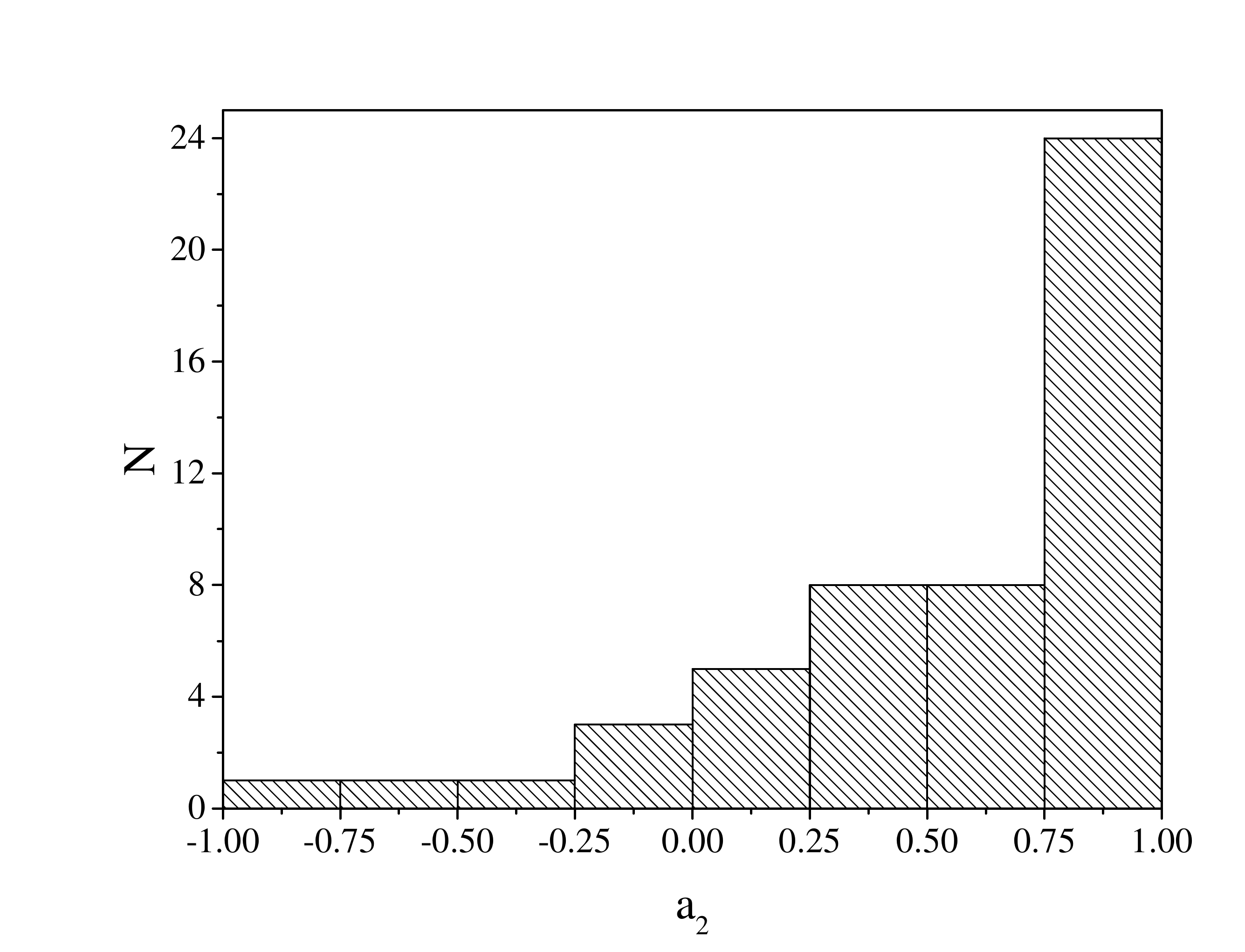}
\includegraphics[bb= 58 11 726 525, clip, width=0.67 \columnwidth]{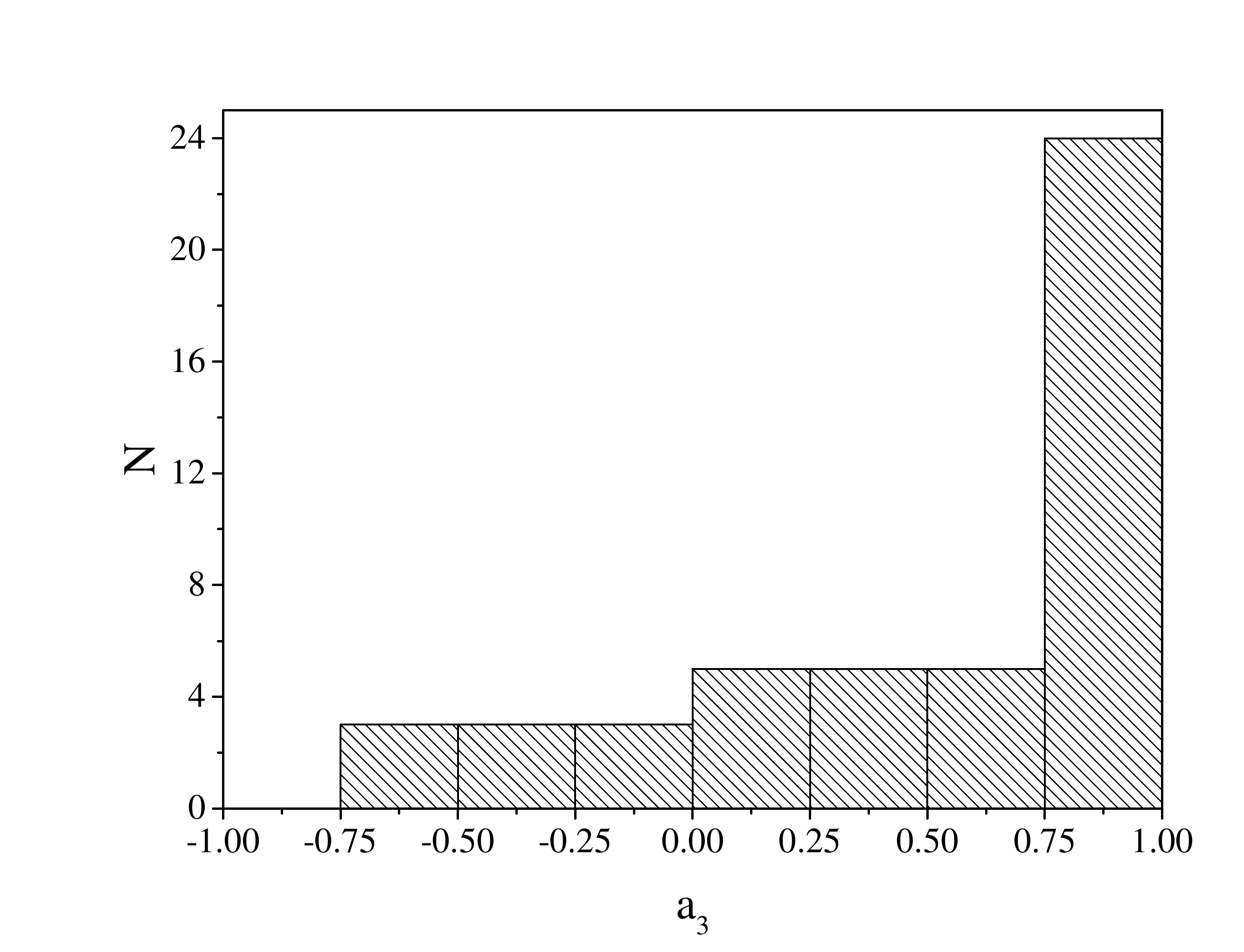}
\caption{Histograms showing the number of objects with a certain spin value $a$ for all individual models: $a_1$ - \citet{raimundo11}, $a_2$ - \citet{trakhtenbrot14}, $a_3$ - \citet{du14}.}
\label{fig06}
\end{figure*}

\section{Results}

\subsection{Results of the first model \citep{raimundo11}}

First we analyse the model from \citet{raimundo11}. The linear fitting of the dependence of the spin on the cosmological redshift for this model (first panel from Fig.~\ref{fig03}) gives us: $a_1 = (0.21 \pm 0.14)\log{z} + (0.71\pm 0.23)$. The dependence of the spin on the bolometric luminosity (second panel from Fig.~\ref{fig03}) gives us: $a_1 = (0.07 \pm 0.09) \log{L_\text{bol}\text{[erg/s]}} - (2.79 \pm 3.86)$. The dependence of the spin on the mass of SMBH (third panel from Fig.\ref{fig03}) gives us: $a_1 = (0.27 \pm 0.11) \log{M_\text{BH}/M_\odot} - (1.65 \pm 0.81)$.

\subsection{Results of the second model \citep{trakhtenbrot14}}

Hear we analyse the model from \citet{trakhtenbrot14}. The linear fitting of the dependence of the spin on the cosmological redshift for this model (first panel from Fig.~\ref{fig04}) gives us: $a_2 = (0.23 \pm 0.11)\log{z} + (0.95\pm 0.19)$ . The dependence of the spin on the bolometric luminosity (second panel from Fig.~\ref{fig04}) gives us: $a_2 = (0.13 \pm 0.07) \log{L_\text{bol}\text{[erg/s]}} - (5.09 \pm 3.05)$. The dependence of the spin on the mass of SMBH (third panel from Fig.\ref{fig04}) gives us: $a_2 = (0.32 \pm 0.08) \log{M_\text{BH}/M_\odot} - (1.88 \pm 0.64)$.

\subsection{Results of the third model \citep{du14}}

And finally we analyse the model from \citet{du14}. The linear fitting of the dependence of the spin on the cosmological redshift for this model (first panel from Fig.~\ref{fig05}) gives us: $a_3 = (0.44 \pm 0.11)\log{z} + (1.21\pm 0.20)$. The dependence of the spin on the bolometric luminosity (second panel from Fig.~\ref{fig05}) gives us: $a_3 = (0.25 \pm 0.07) \log{L_\text{bol}\text{[erg/s]}} - (10.59 \pm 3.21)$. The dependence of the spin on the mass of SMBH (third panel from Fig.\ref{fig05}) gives us: $a_3 = (0.46 \pm 0.09) \log{M_\text{BH}/M_\odot} - (3.00 \pm 0.71)$.

\subsection{Comparison of models and statistical analysis}

The linear fitting of the dependence of the spin on the cosmological redshift for all three models shows the same trend and gives close results within a margin of error. However first \citep{raimundo11} and second \citep{trakhtenbrot14} models have significantly larger errors and their results can rather be considered only as qualitative. We can see that the spin value decreases with cosmic time in all models. This result is in general agreement with results of theoretical calculations for low redshift AGNs. As SMBHs get more massive and galaxies more gas poor, the contribution from binary coalescences to the total BH mass growth increases and tends to decrease the magnitude of spins and change their direction \citep{volonteri13,dubois14,griffin19}.

Concerning the dependence of the spin on the bolometric luminosity,  second \citep{trakhtenbrot14} and third \citep{du14} models show similar trend. The spin increases with increasing bolometric luminosity. But only the results of the third \citep{du14} model can be considered as quantitative. Results of first \citep{raimundo11} model is uncertain due to large errors. This result is quite expected, because as the spin increases, the radiative efficiency also increases \citep{bardeen72,novikov73,krolik07,krolik07b}, which in turn increases luminosity.

Now we consider the dependence of the spin on the mass of SMBH. All models give close results within a margin of error. We can see that the spin value increases with the increasing mass. This result is also in general agreement with the mentioned above theoretical calculations \citep{volonteri13,dubois14,griffin19}.

Fig.~\ref{fig06} presents histograms showing the number of objects with a certain spin value $a$ for all individual models. All histograms show similar distributions. There is a pronounced peak in the distribution in $0.75 < a < 1.0$ range. $\sim$40\% of objects have spin $a > 0.75$ and $\sim$50\% of objects have spin $a > 0.5$. This results are in a good agreement with our previous results \citep{afanasiev18} and with results of other authors \citep{trakhtenbrot14}.

\section{Conclusions}

We estimated the spin value for a sample of 111 AGNs from \citet{marin16} using 3 popular models \citep{raimundo11,du14,trakhtenbrot14} connecting the radiative efficiency (which depends significantly on the spin) with such parameters of AGNs as mass of SMBH $M_\text{BH}$, angle between the line of sight and the axis of the accretion disk $i$ and bolometric luminosity $L_\text{bol}$.

All three models show qualitatively the same trends in relation to all parameters. But only the results of the third model \citep{du14} can be considered as quantitative. Thus, we can conclude that the third model is the most preferable. However, it should be noted that for more accurate conclusions, an analysis of a larger sample of objects is required.

Analysis of the obtained data shown that the spin value increases with the increasing cosmological redshift, i.e. spin value decreases with cosmic time. For low redshift AGNs, as SMBHs get more massive and galaxies more gas poor, the contribution from binary coalescences to the total BH mass growth increases and tends to decrease the magnitude of spins and change their direction \citep{volonteri13,dubois14,griffin19}.

Also we found that the spin value increases with the increasing bolometric luminosity and mass of SMBH. This is the expected result that generally corresponds to theoretical calculations. As the spin increases, the radiative efficiency also increases \citep{bardeen72,novikov73,krolik07,krolik07b}, which in turn increases luminosity.

Analysis of the distribution of the mean spin values shown a pronounced peak in the distribution in $0.75 < a < 1.0$ range. $\sim$40\% of objects have spin $a > 0.75$ and $\sim$50\% of objects have spin $a > 0.5$. This results are in a good agreement with our previous results \citep{afanasiev18} and with the results of other authors \citep{trakhtenbrot14}.

Two serious difficulties in this problem should be noted. First, the presence of several models that give different predictions on the radiative efficiency dependence on other parameters. And, second, the problem of definition of the bolometric correction factors. Both of these problems can be partially solved by analysing the large samples of objects.

\section*{Acknowledgments}

This research was supported by the grant of Russian Science Foundation project number 20-12-00030 ``Investigation of geometry and kinematics of ionized gas in active galactic nuclei by polarimetry methods''.

\bibliography{paper}

\begin{thebibliography}{}

\bibitem [\protect \citeauthoryear {%
{Afanasiev}%
, {Gnedin}%
, {Piotrovich}%
, {Natsvlishvili}%
\BCBL {}\ \BBA {} {Buliga}%
}{%
{Afanasiev}%
\ \protect \BOthers {.}}{%
{\protect \APACyear {2018}}%
}]{%
afanasiev18}
\APACinsertmetastar {%
afanasiev18}%
\begin{APACrefauthors}%
{Afanasiev}, V\BPBI L.%
, {Gnedin}, Y\BPBI N.%
, {Piotrovich}, M\BPBI Y.%
, {Natsvlishvili}, T\BPBI M.%
\BCBL {}\ \BBA {} {Buliga}, S\BPBI D.%
\end{APACrefauthors}%
\unskip\
\newblock
\APACrefYearMonthDay{2018}{}{},
\newblock
\unskip
\newblock
\APACjournalVolNumPages{Astronomy Letters}{44}{}{362-369}.
\PrintBackRefs{\CurrentBib}

\bibitem [\protect \citeauthoryear {%
{Bardeen}%
, {Press}%
\BCBL {}\ \BBA {} {Teukolsky}%
}{%
{Bardeen}%
\ \protect \BOthers {.}}{%
{\protect \APACyear {1972}}%
}]{%
bardeen72}
\APACinsertmetastar {%
bardeen72}%
\begin{APACrefauthors}%
{Bardeen}, J\BPBI M.%
, {Press}, W\BPBI H.%
\BCBL {}\ \BBA {} {Teukolsky}, S\BPBI A.%
\end{APACrefauthors}%
\unskip\
\newblock
\APACrefYearMonthDay{1972}{}{},
\newblock
\unskip
\newblock
\APACjournalVolNumPages{\apj}{178}{}{347-370}.
\PrintBackRefs{\CurrentBib}

\bibitem [\protect \citeauthoryear {%
{Blandford}%
\ \BBA {} {Payne}%
}{%
{Blandford}%
\ \BBA {} {Payne}%
}{%
{\protect \APACyear {1982}}%
}]{%
blandford82}
\APACinsertmetastar {%
blandford82}%
\begin{APACrefauthors}%
{Blandford}, R\BPBI D.%
\BCBT {}\ \BBA {} {Payne}, D\BPBI G.%
\end{APACrefauthors}%
\unskip\
\newblock
\APACrefYearMonthDay{1982}{}{},
\newblock
\unskip
\newblock
\APACjournalVolNumPages{\mnras}{199}{}{883-903}.
\PrintBackRefs{\CurrentBib}

\bibitem [\protect \citeauthoryear {%
{Blandford}%
\ \BBA {} {Znajek}%
}{%
{Blandford}%
\ \BBA {} {Znajek}%
}{%
{\protect \APACyear {1977}}%
}]{%
blandford77}
\APACinsertmetastar {%
blandford77}%
\begin{APACrefauthors}%
{Blandford}, R\BPBI D.%
\BCBT {}\ \BBA {} {Znajek}, R\BPBI L.%
\end{APACrefauthors}%
\unskip\
\newblock
\APACrefYearMonthDay{1977}{}{},
\newblock
\unskip
\newblock
\APACjournalVolNumPages{\mnras}{179}{}{433-456}.
\PrintBackRefs{\CurrentBib}

\bibitem [\protect \citeauthoryear {%
{Cheng}%
\ \protect \BOthers {.}}{%
{Cheng}%
\ \protect \BOthers {.}}{%
{\protect \APACyear {2019}}%
}]{%
cheng19}
\APACinsertmetastar {%
cheng19}%
\begin{APACrefauthors}%
{Cheng}, H.%
, {Yuan}, W.%
, {Liu}, H\BHBI Y.%
, {Breeveld}, A\BPBI A.%
, {Jin}, C.%
\BCBL {}\ \BBA {} {Liu}, B.%
\end{APACrefauthors}%
\unskip\
\newblock
\APACrefYearMonthDay{2019}{}{},
\newblock
\unskip
\newblock
\APACjournalVolNumPages{\mnras}{487}{3}{3884-3903}.
\PrintBackRefs{\CurrentBib}

\bibitem [\protect \citeauthoryear {%
{Daly}%
}{%
{Daly}%
}{%
{\protect \APACyear {2011}}%
}]{%
daly11}
\APACinsertmetastar {%
daly11}%
\begin{APACrefauthors}%
{Daly}, R\BPBI A.%
\end{APACrefauthors}%
\unskip\
\newblock
\APACrefYearMonthDay{2011}{}{},
\newblock
\unskip
\newblock
\APACjournalVolNumPages{\mnras}{414}{}{1253-1262}.
\PrintBackRefs{\CurrentBib}

\bibitem [\protect \citeauthoryear {%
{Davis}%
\ \BBA {} {Laor}%
}{%
{Davis}%
\ \BBA {} {Laor}%
}{%
{\protect \APACyear {2011}}%
}]{%
davis11}
\APACinsertmetastar {%
davis11}%
\begin{APACrefauthors}%
{Davis}, S\BPBI W.%
\BCBT {}\ \BBA {} {Laor}, A.%
\end{APACrefauthors}%
\unskip\
\newblock
\APACrefYearMonthDay{2011}{}{},
\newblock
\unskip
\newblock
\APACjournalVolNumPages{\apj}{728}{}{98}.
\PrintBackRefs{\CurrentBib}

\bibitem [\protect \citeauthoryear {%
{Du}%
\ \protect \BOthers {.}}{%
{Du}%
\ \protect \BOthers {.}}{%
{\protect \APACyear {2014}}%
}]{%
du14}
\APACinsertmetastar {%
du14}%
\begin{APACrefauthors}%
{Du}, P.%
, {Hu}, C.%
, {Lu}, K\BHBI X.%
\ et al.\end{APACrefauthors}%
\unskip\
\newblock
\APACrefYearMonthDay{2014}{}{},
\newblock
\unskip
\newblock
\APACjournalVolNumPages{\apj}{782}{}{45}.
\PrintBackRefs{\CurrentBib}

\bibitem [\protect \citeauthoryear {%
{Dubois}%
, {Volonteri}%
\BCBL {}\ \BBA {} {Silk}%
}{%
{Dubois}%
\ \protect \BOthers {.}}{%
{\protect \APACyear {2014}}%
}]{%
dubois14}
\APACinsertmetastar {%
dubois14}%
\begin{APACrefauthors}%
{Dubois}, Y.%
, {Volonteri}, M.%
\BCBL {}\ \BBA {} {Silk}, J.%
\end{APACrefauthors}%
\unskip\
\newblock
\APACrefYearMonthDay{2014}{}{},
\newblock
\unskip
\newblock
\APACjournalVolNumPages{\mnras}{440}{2}{1590-1606}.
\PrintBackRefs{\CurrentBib}

\bibitem [\protect \citeauthoryear {%
{Duras}%
\ \protect \BOthers {.}}{%
{Duras}%
\ \protect \BOthers {.}}{%
{\protect \APACyear {2020}}%
}]{%
duras20}
\APACinsertmetastar {%
duras20}%
\begin{APACrefauthors}%
{Duras}, F.%
, {Bongiorno}, A.%
, {Ricci}, F.%
\ et al.\end{APACrefauthors}%
\unskip\
\newblock
\APACrefYearMonthDay{2020}{}{},
\newblock
\unskip
\newblock
\APACjournalVolNumPages{\aap}{636}{}{A73}.
\PrintBackRefs{\CurrentBib}

\bibitem [\protect \citeauthoryear {%
{Garofalo}%
, {Evans}%
\BCBL {}\ \BBA {} {Sambruna}%
}{%
{Garofalo}%
\ \protect \BOthers {.}}{%
{\protect \APACyear {2010}}%
}]{%
garofalo10}
\APACinsertmetastar {%
garofalo10}%
\begin{APACrefauthors}%
{Garofalo}, D.%
, {Evans}, D\BPBI A.%
\BCBL {}\ \BBA {} {Sambruna}, R\BPBI M.%
\end{APACrefauthors}%
\unskip\
\newblock
\APACrefYearMonthDay{2010}{}{},
\newblock
\unskip
\newblock
\APACjournalVolNumPages{\mnras}{406}{}{975-986}.
\PrintBackRefs{\CurrentBib}

\bibitem [\protect \citeauthoryear {%
{Griffin}%
\ \protect \BOthers {.}}{%
{Griffin}%
\ \protect \BOthers {.}}{%
{\protect \APACyear {2019}}%
}]{%
griffin19}
\APACinsertmetastar {%
griffin19}%
\begin{APACrefauthors}%
{Griffin}, A\BPBI J.%
, {Lacey}, C\BPBI G.%
, {Gonzalez-Perez}, V.%
, {Lagos}, C\BPBI d\BPBI P.%
, {Baugh}, C\BPBI M.%
\BCBL {}\ \BBA {} {Fanidakis}, N.%
\end{APACrefauthors}%
\unskip\
\newblock
\APACrefYearMonthDay{2019}{}{},
\newblock
\unskip
\newblock
\APACjournalVolNumPages{\mnras}{487}{1}{198-227}.
\PrintBackRefs{\CurrentBib}

\bibitem [\protect \citeauthoryear {%
{Hopkins}%
, {Richards}%
\BCBL {}\ \BBA {} {Hernquist}%
}{%
{Hopkins}%
\ \protect \BOthers {.}}{%
{\protect \APACyear {2007}}%
}]{%
hopkins07}
\APACinsertmetastar {%
hopkins07}%
\begin{APACrefauthors}%
{Hopkins}, P\BPBI F.%
, {Richards}, G\BPBI T.%
\BCBL {}\ \BBA {} {Hernquist}, L.%
\end{APACrefauthors}%
\unskip\
\newblock
\APACrefYearMonthDay{2007}{}{},
\newblock
\unskip
\newblock
\APACjournalVolNumPages{\apj}{654}{2}{731-753}.
\PrintBackRefs{\CurrentBib}

\bibitem [\protect \citeauthoryear {%
{Krolik}%
}{%
{Krolik}%
}{%
{\protect \APACyear {2007}}%
}]{%
krolik07}
\APACinsertmetastar {%
krolik07}%
\begin{APACrefauthors}%
{Krolik}, J\BPBI H.%
\end{APACrefauthors}%
\unskip\
\newblock
\APACrefYearMonthDay{2007}{}{},
\newblock
{\BBOQ}\APACrefatitle {{Making black holes visible: accretion, radiation, and
  jets}} {{Making black holes visible: accretion, radiation, and jets}}.{\BBCQ}
\newblock
\BIn{} \APACrefbtitle {{2007 STScI Spring Symposium on Black Holes}} {{2007
  STScI Spring Symposium on Black Holes}}\ \BPG~309-321.
\PrintBackRefs{\CurrentBib}

\bibitem [\protect \citeauthoryear {%
{Krolik}%
, {Hawley}%
\BCBL {}\ \BBA {} {Hirose}%
}{%
{Krolik}%
\ \protect \BOthers {.}}{%
{\protect \APACyear {2007}}%
}]{%
krolik07b}
\APACinsertmetastar {%
krolik07b}%
\begin{APACrefauthors}%
{Krolik}, J\BPBI H.%
, {Hawley}, J\BPBI F.%
\BCBL {}\ \BBA {} {Hirose}, S.%
\end{APACrefauthors}%
\unskip\
\newblock
\APACrefYearMonthDay{2007}{}{},
\newblock
{\BBOQ}\APACrefatitle {{The Relationship between Accretion Disks and Jets}}
  {{The Relationship between Accretion Disks and Jets}}.{\BBCQ}
\newblock
\BIn{} \APACrefbtitle {Revista Mexicana de Astronomia y Astrofisica} {Revista
  Mexicana de Astronomia y Astrofisica}\ \BVOL~27, \BPG~1-7.
\PrintBackRefs{\CurrentBib}

\bibitem [\protect \citeauthoryear {%
{Lawther}%
, {Vestergaard}%
, {Raimundo}%
\BCBL {}\ \BBA {} {Grupe}%
}{%
{Lawther}%
\ \protect \BOthers {.}}{%
{\protect \APACyear {2017}}%
}]{%
lawther17}
\APACinsertmetastar {%
lawther17}%
\begin{APACrefauthors}%
{Lawther}, D.%
, {Vestergaard}, M.%
, {Raimundo}, S.%
\BCBL {}\ \BBA {} {Grupe}, D.%
\end{APACrefauthors}%
\unskip\
\newblock
\APACrefYearMonthDay{2017}{}{},
\newblock
\unskip
\newblock
\APACjournalVolNumPages{\mnras}{467}{4}{4674-4710}.
\PrintBackRefs{\CurrentBib}

\bibitem [\protect \citeauthoryear {%
{Marin}%
}{%
{Marin}%
}{%
{\protect \APACyear {2016}}%
}]{%
marin16}
\APACinsertmetastar {%
marin16}%
\begin{APACrefauthors}%
{Marin}, F.%
\end{APACrefauthors}%
\unskip\
\newblock
\APACrefYearMonthDay{2016}{}{},
\newblock
\unskip
\newblock
\APACjournalVolNumPages{\mnras}{460}{}{3679-3705}.
\PrintBackRefs{\CurrentBib}

\bibitem [\protect \citeauthoryear {%
{Netzer}%
}{%
{Netzer}%
}{%
{\protect \APACyear {2019}}%
}]{%
netzer19}
\APACinsertmetastar {%
netzer19}%
\begin{APACrefauthors}%
{Netzer}, H.%
\end{APACrefauthors}%
\unskip\
\newblock
\APACrefYearMonthDay{2019}{}{},
\newblock
\unskip
\newblock
\APACjournalVolNumPages{\mnras}{488}{4}{5185-5191}.
\PrintBackRefs{\CurrentBib}

\bibitem [\protect \citeauthoryear {%
{Novikov}%
\ \BBA {} {Thorne}%
}{%
{Novikov}%
\ \BBA {} {Thorne}%
}{%
{\protect \APACyear {1973}}%
}]{%
novikov73}
\APACinsertmetastar {%
novikov73}%
\begin{APACrefauthors}%
{Novikov}, I\BPBI D.%
\BCBT {}\ \BBA {} {Thorne}, K\BPBI S.%
\end{APACrefauthors}%
\unskip\
\newblock
\APACrefYearMonthDay{1973}{}{},
\newblock
{\BBOQ}\APACrefatitle {{Astrophysics of black holes.}} {{Astrophysics of black
  holes.}}{\BBCQ}
\newblock
\BIn{} C.~{Dewitt}\ \BBA {} B\BPBI S.~{Dewitt}\ (\BEDS), \APACrefbtitle {Black
  Holes (Les Astres Occlus)} {Black Holes (Les Astres Occlus)}\ \BPG~343-450.
\newblock
\APACaddressPublisher{New York}{Gordon and Breach}.
\PrintBackRefs{\CurrentBib}

\bibitem [\protect \citeauthoryear {%
{Raimundo}%
, {Fabian}%
, {Vasudevan}%
, {Gandhi}%
\BCBL {}\ \BBA {} {Wu}%
}{%
{Raimundo}%
\ \protect \BOthers {.}}{%
{\protect \APACyear {2012}}%
}]{%
raimundo11}
\APACinsertmetastar {%
raimundo11}%
\begin{APACrefauthors}%
{Raimundo}, S\BPBI I.%
, {Fabian}, A\BPBI C.%
, {Vasudevan}, R\BPBI V.%
, {Gandhi}, P.%
\BCBL {}\ \BBA {} {Wu}, J.%
\end{APACrefauthors}%
\unskip\
\newblock
\APACrefYearMonthDay{2012}{}{},
\newblock
\unskip
\newblock
\APACjournalVolNumPages{\mnras}{419}{}{2529-2544}.
\PrintBackRefs{\CurrentBib}

\bibitem [\protect \citeauthoryear {%
{Richards}%
\ \protect \BOthers {.}}{%
{Richards}%
\ \protect \BOthers {.}}{%
{\protect \APACyear {2006}}%
}]{%
richards06}
\APACinsertmetastar {%
richards06}%
\begin{APACrefauthors}%
{Richards}, G\BPBI T.%
, {Lacy}, M.%
, {Storrie-Lombardi}, L\BPBI J.%
\ et al.\end{APACrefauthors}%
\unskip\
\newblock
\APACrefYearMonthDay{2006}{}{},
\newblock
\unskip
\newblock
\APACjournalVolNumPages{\apjs}{166}{}{470-497}.
\PrintBackRefs{\CurrentBib}

\bibitem [\protect \citeauthoryear {%
{Shakura}%
\ \BBA {} {Sunyaev}%
}{%
{Shakura}%
\ \BBA {} {Sunyaev}%
}{%
{\protect \APACyear {1973}}%
}]{%
shakura73}
\APACinsertmetastar {%
shakura73}%
\begin{APACrefauthors}%
{Shakura}, N\BPBI I.%
\BCBT {}\ \BBA {} {Sunyaev}, R\BPBI A.%
\end{APACrefauthors}%
\unskip\
\newblock
\APACrefYearMonthDay{1973}{}{},
\newblock
\unskip
\newblock
\APACjournalVolNumPages{\aap}{24}{}{337-355}.
\PrintBackRefs{\CurrentBib}

\bibitem [\protect \citeauthoryear {%
{Thorne}%
}{%
{Thorne}%
}{%
{\protect \APACyear {1974}}%
}]{%
thorne74}
\APACinsertmetastar {%
thorne74}%
\begin{APACrefauthors}%
{Thorne}, K\BPBI S.%
\end{APACrefauthors}%
\unskip\
\newblock
\APACrefYearMonthDay{1974}{}{},
\newblock
\unskip
\newblock
\APACjournalVolNumPages{\apj}{191}{}{507-520}.
\PrintBackRefs{\CurrentBib}

\bibitem [\protect \citeauthoryear {%
{Trakhtenbrot}%
}{%
{Trakhtenbrot}%
}{%
{\protect \APACyear {2014}}%
}]{%
trakhtenbrot14}
\APACinsertmetastar {%
trakhtenbrot14}%
\begin{APACrefauthors}%
{Trakhtenbrot}, B.%
\end{APACrefauthors}%
\unskip\
\newblock
\APACrefYearMonthDay{2014}{}{},
\newblock
\unskip
\newblock
\APACjournalVolNumPages{\apjl}{789}{}{L9}.
\PrintBackRefs{\CurrentBib}

\bibitem [\protect \citeauthoryear {%
{Volonteri}%
, {Sikora}%
, {Lasota}%
\BCBL {}\ \BBA {} {Merloni}%
}{%
{Volonteri}%
\ \protect \BOthers {.}}{%
{\protect \APACyear {2013}}%
}]{%
volonteri13}
\APACinsertmetastar {%
volonteri13}%
\begin{APACrefauthors}%
{Volonteri}, M.%
, {Sikora}, M.%
, {Lasota}, J\BPBI P.%
\BCBL {}\ \BBA {} {Merloni}, A.%
\end{APACrefauthors}%
\unskip\
\newblock
\APACrefYearMonthDay{2013}{}{},
\newblock
\unskip
\newblock
\APACjournalVolNumPages{\apj}{775}{2}{94}.
\PrintBackRefs{\CurrentBib}

\end{thebibliography}

\end{document}